\documentclass[preprint]{aastex62}

\usepackage{natbib}
\usepackage{CJK}
\usepackage{color}
\usepackage{graphicx}

\newcommand{\LFunit}{counts\,s$^{-1}$\,deg$^{-2}$}
\newcommand{\RCS}{$\chi^2/\textrm{dof}$ }
\newcommand{\fluxunit}{photons\,s$^{-1}$\,cm$^{-2}$}
\newcommand{\kTe}{kT_{\rm e}}
\newcommand{\kTz}{kT_{\rm z}}
\newcommand{\Msun}{M_\odot}

\submitjournal{ApJ}

\begin{document}
\begin{CJK}{Bg5}{bsmi}
	
	\title{{An XMM-Newton X-ray View of Supernova Remnant W49B: Revisiting its Recombining Plasmas and Progenitor Type}}
	
	\correspondingauthor{Yang Chen}
	\email{ygchen@nju.edu.cn}
	
	\author{Lei Sun (®]½U)}
	\affiliation{Department of Astronomy, Nanjing University, Nanjing 210023, China}
	
	\author{Yang Chen (³¯¶§)}
	\affiliation{Department of Astronomy, Nanjing University, Nanjing 210023, China}
	\affiliation{Key Laboratory of Modern Astronomy and Astrophysics, Nanjing University, Ministry of Education, China}

	\begin{abstract}
	We present a comprehensive X-ray spectroscopy and imaging study of supernova remnant W49B using archival {\it XMM-Newton} observations. The overionization state of the shocked ejecta in W49B is clearly indicated by the radiative recombination continua of Si XIV, S XV, and Fe XXV, combined with the Ly$\alpha$ lines of Ca and Fe. The line flux images of W49B indicate high emission measures of the central bar-like region for almost all the emission lines, while the equivalent width maps reveal a stratified structure for the metal abundance distributions. The global spectrum of W49B is well reproduced by a model containing one collisional ionization equilibrium (CIE) plasma component and two recombining plasma (RP) components. The CIE plasma represents the shocked interstellar medium, which dominates the X-ray emitting volume in W49B with a mass $\sim450\Msun$. The two RP components with a total mass $\sim4.6\Msun$ are both dominated by the ejecta material, but characterized by different electron temperatures ($\sim1.60$\,keV and $\sim0.64$\,keV). The recombination ages of the RP components are estimated as $\sim6000$\,yr and $\sim3400$\,yr, respectively. We then reveal the possibility of a thermal conduction origin for the high-temperature RP in W49B by calculating the conduction timescale. The metal abundance ratios of the ejecta in W49B are roughly consistent with a core-collapse explosion model with a $\lesssim15\Msun$ progenitor, except for a rather high Mn/Fe. A Type Ia origin can explain the Mn abundance, while it predicts much higher ejecta masses than observed values for all the metal species considered in our analysis. 
	\end{abstract}
	
	\section{Introduction} \label{sec:intro}
	With the advances in X-ray imaging spectroscopy over the last {two decades}, substantial progress {has} been made in the study of supernova remnant (SNR) {physics}. Two of the major {concerns have been intensively given to} the evolution mechanism of SNR {plasma} and {its} implications for supernova (SN) physics.
		
		\subsection{Recombining plasma}
		In the early phases of the SNR evolution, the ambient materials and the SN ejecta are compressed and heated by the forward and reverse shocks, {respectively,} and {leaving behind X-ray emitting plasmas.} Due to the rather low density, the shock-heated plasma is expected to be in the non-equilibrium ionization (NEI) state for a long time. Most of the young or middle-aged SNRs { are} reported to involve ionizing (underionized) plasma (IP), where the ionization temperature ($\kTz$) is still lower than the electron temperature ($\kTe$). However, recent X-ray spectroscopic studies of several SNRs \citep[e.g., IC 443, G359.1-0.5, W28, W44, etc.:][]{2002ApJ...572..897K,2009ApJ...705L...6Y,2011PASJ...63..527O,2012PASJ...64...81S,2012PASJ...64..141U} revealed the existence of recombining (overionized) plasma (RP), where $\kTz$ goes even higher than $\kTe$. So far, RP has been found in more than a dozen of SNRs, which may represent a new subclass of SNRs.
		
		{The physical origin of the RPs observed in SNRs is still under debate.} 
		Theoretically, there are two possible ways to produce overionized plasma: increase of $\kTz$ (extra-ionization) or decrease of $\kTe$ (electron cooling). The extra-ionization process can be caused by {suprathermal} electrons \citep[e.g.,][]{2011PASJ...63..527O} or high-energy photons \citep[e.g.,][]{2002ApJ...572..897K}. On the other hand, the electron cooling scenario, which can better apply to the SNR evolution, may arise from adiabatic expansion \citep[e.g.,][]{1989MNRAS.236..885I} {and/}or thermal conduction \citep[e.g.,][]{2002ApJ...572..897K,2011MNRAS.415..244Z}. In addition, {various} scenarios (such as the adiabatic expansion and the thermal conduction scenario) may simultaneously contribute to the formation of RP, as indicated by \citet{2011MNRAS.415..244Z} and \citet{2019ApJ...875...81Z} based on hydrodynamic simulations. 
		
		While the nature of RP remains unclear, {spatially-resolved X-ray spectroscopic studies of SNRs have provided us with significant clues.}
		A higher degree of overionization (or a lower $\kTe$) towards the cold dense interstellar medium (ISM) may {be indicative of} thermal conduction scenario \citep[e.g., G166.0+4.2, CTB 1:][]{2017PASJ...69...30M,2018PASJ...70..110K}. In contrast, a higher degree of overionization towards low density regions may {be indicative of} adiabatic expansion scenario \citep[e.g., W49B:][]{2010AA...514L...2M,2013ApJ...777..145L,2018ApJ...868L..35Y}. On the other hand, some common features shared by the RP-detected SNRs may give further implications. {In Table \ref{tab:list} we summarize the basic information and the bulk RP properties of the 15 RP-detected Galactic SNRs. We note that, most of the RP-detected SNRs (a) belong to the group of so-called thermal composite ({or} mixed-morphology) SNRs \citep{2015ApJ...799..103Z}, (b) are interacting with molecular clouds \citep{2010ApJ...712.1147J}, and (c) emit GeV/TeV $\gamma$-rays \citep{2018PASJ...70...75S}.
		
		\subsection{SN progenitor}
		{One of the most important issues regarding SNRs science is the nature of their progenitors and their intrinsic properties.}
		The shock-heated SN ejecta become luminous in the SNR phase, which can probe the progenitor type and the detailed explosion mechanism. Meanwhile, the shocked circumstellar material (CSM) can help to trace the history of the progenitor evolution up to thousands even millions of years before the explosion \citep[see, e.g.,][for a recent review]{2017hsn..book.2233P}. However, identification of the progenitor and explosion type for individual SNRs remains a major challenge in SNR {studies}.
		
		Thermonuclear (Type Ia) SNe typically produce a large among of iron group elements (IGEs) such as Fe and Ni, while core-collapse (CC) SNe usually result in high yields of O, Ne and Mg. Therefore, metal abundances of the ejecta material are commonly used as indicators of the progenitor type. Particularly, detailed analyses on the abundance ratios of different metal species may help to disentangle the ejecta from the CSM, and provide further constraints on the progenitor \citep[e.g.,][]{2019ApJ...872...45S}. Moreover, the progenitor type can also be identified based on the bulk properties of SNRs. The Fe K line centroids detected in {Type} Ia SNRs have been found to be generally lower than those detected in CC SNRs, which may be {an useful} diagnostic to discriminate the SNR type \citep{2014ApJ...785L..27Y,2015ApJ...803..101P}. In addition, \citet{2013ApJ...766...44Y} proposed the Cr-to-Fe equivalent width (EW) ratio ($\gamma_{\rm Cr/Fe}\equiv{\rm EW}_{\rm Cr}/{\rm EW}_{\rm Fe}$) as a discriminant of the progenitor type.
		On the other hand, the X-ray morphology of the SNRs can also be used to constrain their progenitor types \citep{2009ApJ...706L.106L,2011ApJ...732..114L}.
		
		\subsection{SNR W49B}
		SNR W49B is one of the most intriguing objects in SNR {science}. It is a mixed-morphology SNR \citep{1998ApJ...503L.167R} at a distance of 8--11.3\,kpc \citep[e.g.,][]{1972ApJS...24...49R,1994ApJ...437..705M,2014IAUS..296..170C,2014ApJ...793...95Z,2018AJ....155..204R}, and is among the most luminous Galactic SNRs in 1\,GHz radio band, Fe K line X-rays, and GeV $\gamma$-rays \citep[e.g.,][]{1994ApJ...437..705M,2014ApJ...785L..27Y,2018A&A...612A...5H}. W49B is suggested to originate inside a wind-blown bubble interior to a dense molecular cloud \citep{2007ApJ...654..938K}. The CO observations further {reveals the evolution of the SNR in a} molecular cavity at a distance of 9.3\,kpc \citep{2014IAUS..296..170C,2014ApJ...793...95Z}. 
		
		W49B is one of the first RP-detected SNRs \citep{2005ApJ...631..935K,2009ApJ...706L..71O}, represented by the remarkable radiative recombining continuum (RRC) of He-like Fe. The physical origin of the RP has not be fully understood. Spatially resolved X-ray studies have shown that the distribution of the RP is most consistent with the adiabatic cooling scenario \citep{2010AA...514L...2M,2013ApJ...777..145L,2018ApJ...868L..35Y}, while the thermal conduction between plasma and ambient {cold} clouds could also be important based on hydrodynamic simulations \citep{2011MNRAS.415..244Z,2019ApJ...875...81Z}. Notably, W49B {might} be the youngest one \citep[1000--6000\,yr, e.g.,][]{1984MNRAS.207..649P,1985ApJ...296..469S,2000ApJ...532..970H,2018AA...615A.150Z} and the only one originate from Type Ia SN \citep{2018AA...615A.150Z} among the RP-detected SNRs (see Table \ref{tab:list}).
		
		{However, the progenitor SN type of W49B is a perplexing question}
		which has been debated for a long time. It is usually considered as a CC SNR, given its peculiar morphology and ambient environment \citep[e.g.,][]{2006A&A...453..567M,2007ApJ...654..938K}. Moreover, it has been suggested that W49B comes from a jet-driven Type Ib/Ic explosion, and might be the first candidate of its kind which harbors a newborn black hole \citep{2013ApJ...764...50L}. However, recent studies on the X-ray spectrum of W49B shown that the metal abundances of the ejecta are better described by Type Ia SN models \citep{2018AA...615A.150Z}. 
		
		{Aiming at a new insight of the relevant physics of W49B}, we present {an} analysis of the archival {\it XMM-Newton} data of {the SNR}. The paper is organized as follows: Section \ref{sec:data} describes the observation data and the reduction procedure; Section \ref{sec:X-ray} presents the main results, in terms of the global spectra (Section \ref{sec:spec_fit}) and the imaging analysis (Section \ref{sec:EW}); Section \ref{sec:disc} discusses the results, mainly focuses on the new constraints of the RP properties and the SN progenitor; and finally Section \ref{sec:sum} gives a brief summary.
		{The errors quoted in this paper represent 90\% confidence ranges, unless otherwise stated.}
	}
	\section{Observations and Data Reduction} \label{sec:data}
	
	{SNR} W49B was observed with {\it XMM-Newton} EPIC-MOS and EPIC-pn cameras in 2004 (PI: A. Decourchelle) and 2014 (PI: L. A. Lopez). We reduce the data based on {\it XMM-Newton} Science Analysis Software (SAS, version 16.1.0)\footnote{https://www.cosmos.esa.int/web/xmm-newton/sas}. All the observation data files are reprocessed using SAS tasks {\tt emchain} and {\tt epchain}. Then, {\tt mos-filter} and {\tt pn-filter} are used to {filter out} soft proton (SP) flares and to remove affected time intervals. The observation {with ObsID} 0084100601 is heavily contaminated by SP flares, {and} thus is excluded from our analysis. The total good time intervals (GTIs) after SP flare removal are 174.0\,ks, 182.6\,ks and 117.5\,ks for {the} MOS1, MOS2 and pn {data}, respectively, as summarized in Table \ref{tab:obs}.
	
	We use SAS tasks {\tt mos-spectra} and {\tt pn-spectra} to create the images and spectra of a given region in certain energy band. Meanwhile, {\tt mos\_back} and {\tt pn\_back} are used to estimate the quiescent particle background (QPB). For image analysis, we adopt tasks {\tt merge\_comp\_xmm} to produce the combined count images, exposure maps and QPB images of all the observations. We use {\tt adapt\_merge} to adaptively smooth the QPB-subtracted and vignetting-corrected count images. All the spectra are grouped with a minimum of 25 counts per channel using FTOOLS task {\tt grppha}. {XSPEC} (version 12.10.1)\footnote{https://heasarc.gsfc.nasa.gov/xanadu/xspec/} with AtomDB 3.0.9\footnote{http://www.atomdb.org/} is used for spectral analysis.
	
	\section{X-ray Properties} \label{sec:X-ray}
	
	Figure \ref{fig:RGB_whole} shows the merged {\it XMM-Newton} {EPIC} image of W49B, with red {color} for 2.35--2.7\,keV (including primarily the S He$\alpha$ and S Ly$\alpha$), green for 4.4--6.2\,keV (the continuum) and blue for 6.45--6.9\,keV (the Fe K complex). 
	
	\subsection{Global spectra} \label{sec:spec_fit}
	
\begin{figure*}
		\plotone{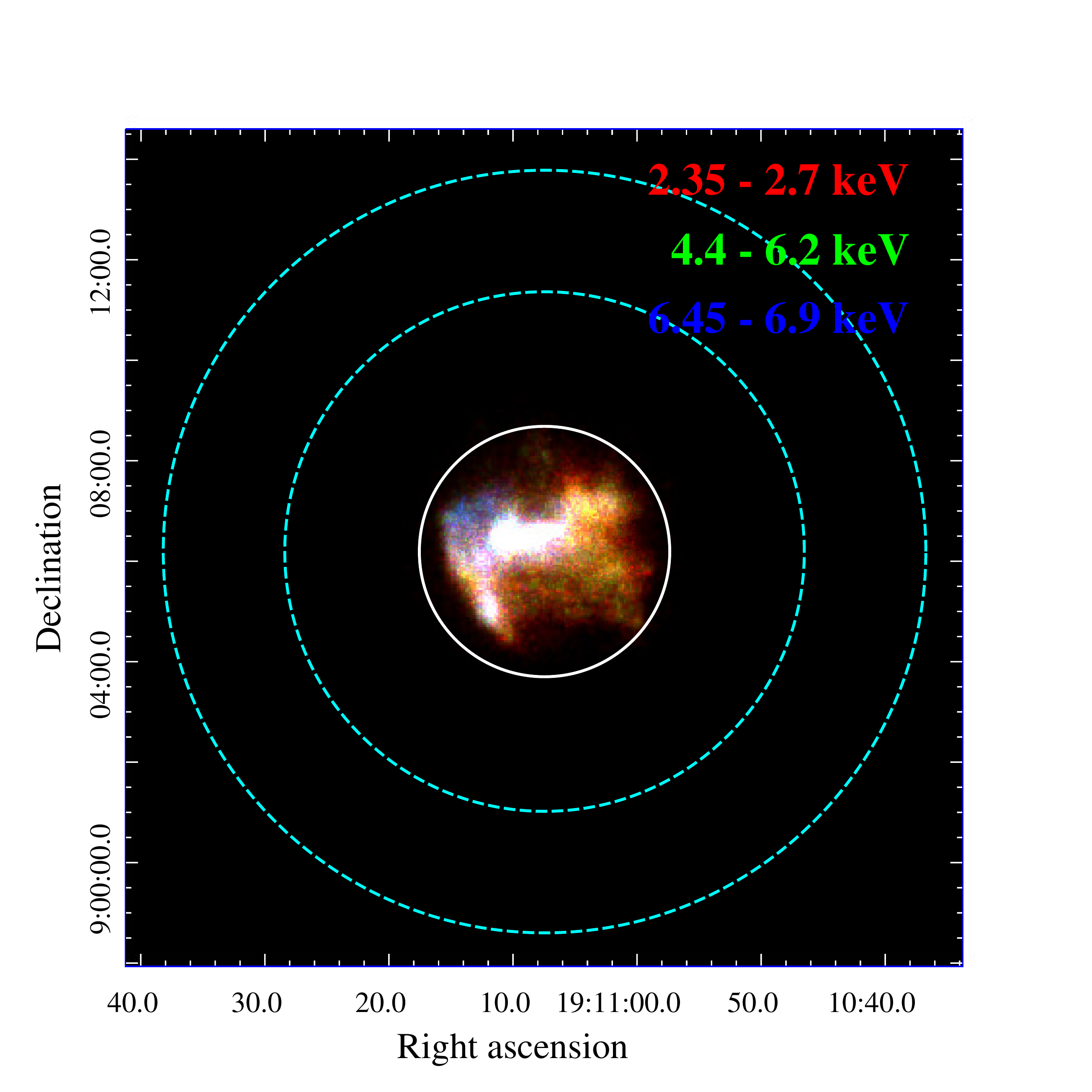}
		\caption{Merged image of the {\it XMM-Newton} EPIC observations of W49B. Red: 2.35--2.7\,keV (S He$\alpha$ and Ly$\alpha$); green: 4.4--6.2\,keV (continuum); blue: 6.45--6.9\,keV (Fe K complex). The white solid circle indicates the region for global spectra extraction and the cyan dashed annulus indicates the region for background spectra extraction. \label{fig:RGB_whole}}
	\end{figure*}

\begin{figure*}
		\plottwo{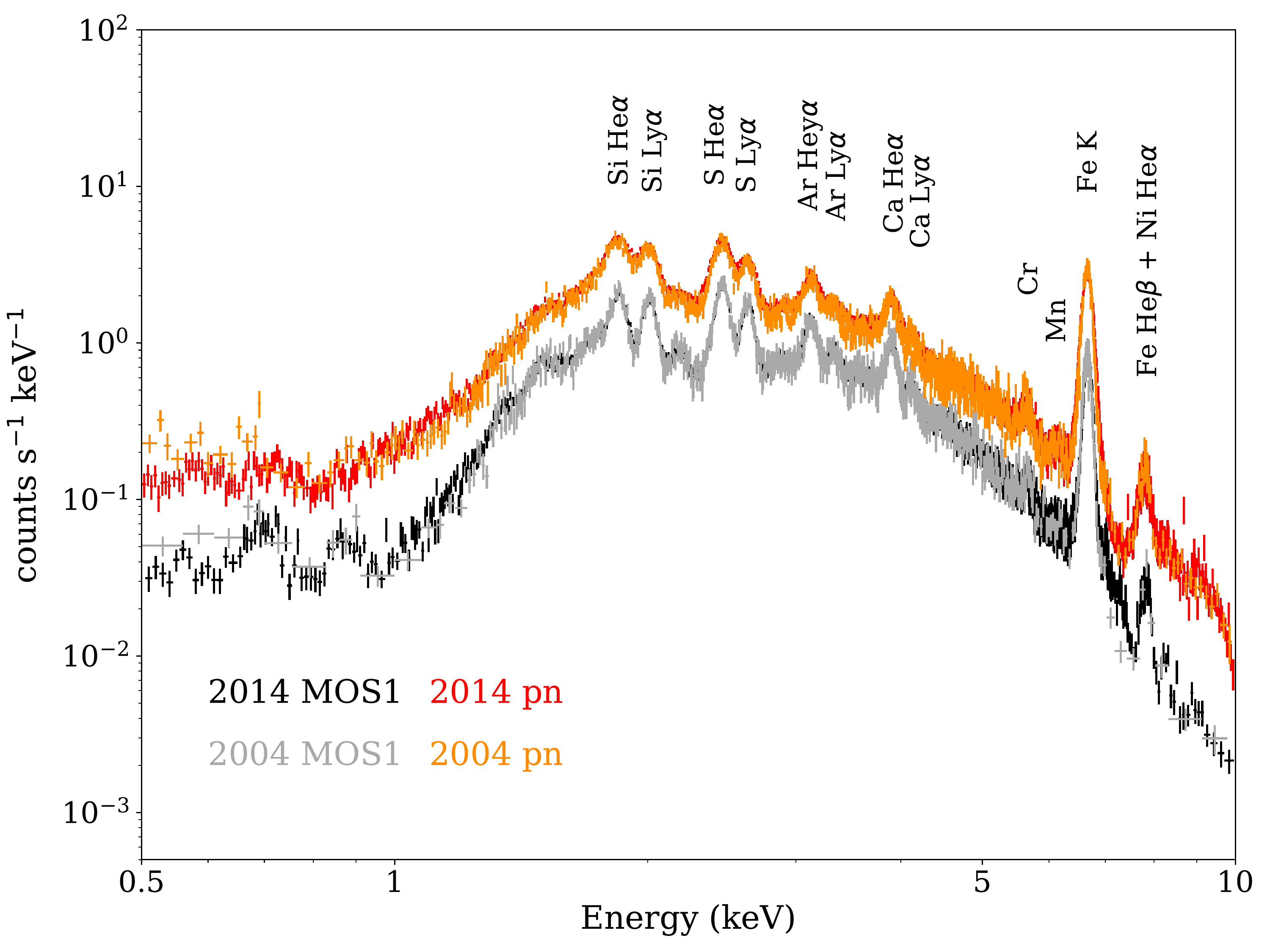}{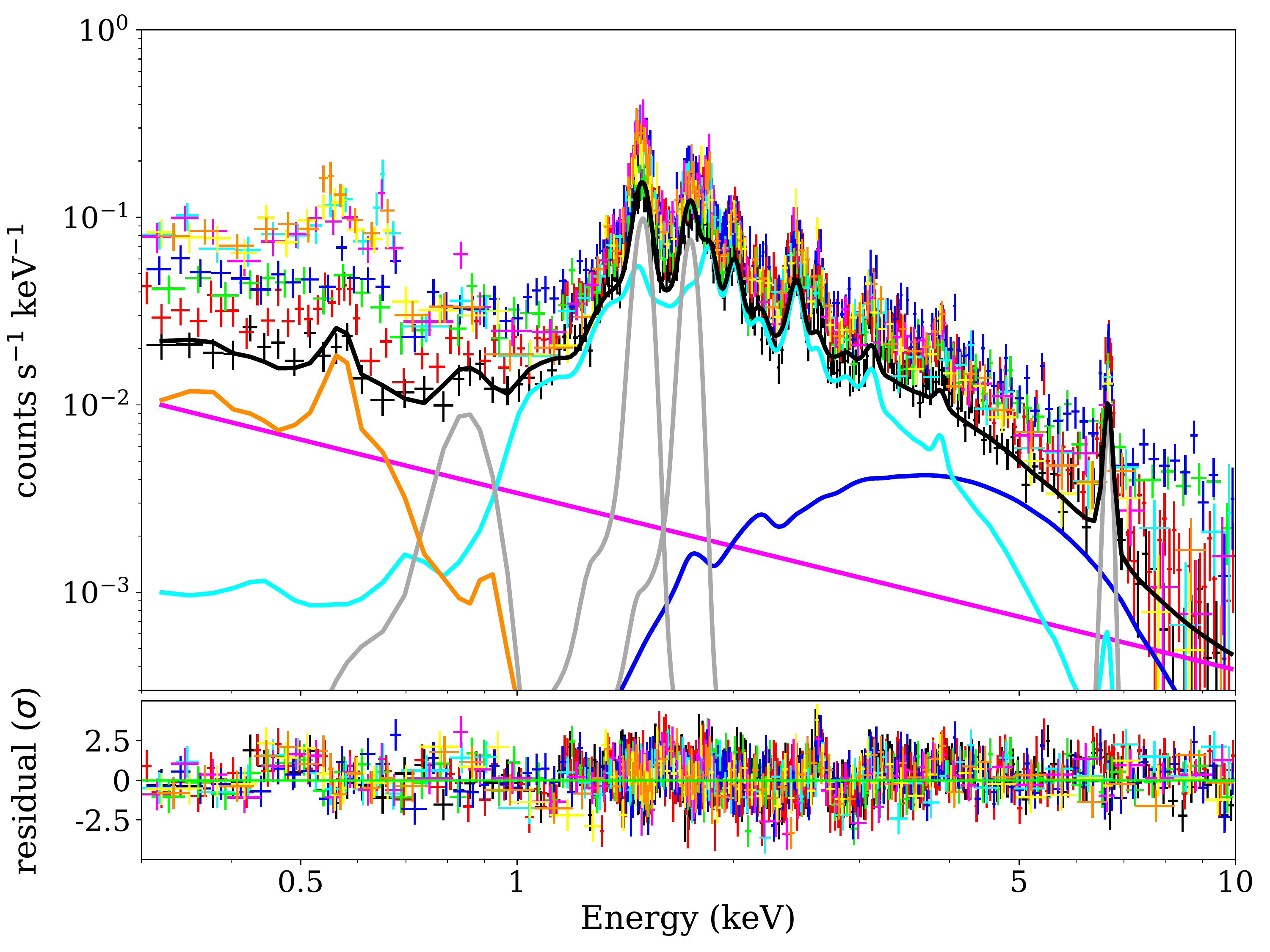}
		\caption{Left: a comparison between the global spectra of W49B taken in 2014 and those taken in 2004. The spectra are extracted from the region indicated by the white circle in Figure \ref{fig:RGB_whole}. Right: {the background spectra (data points), with the best-fit model ({solid curves, different colors for different components: CXB, LHB, Galactic emission, SP contamination, and the Gaussian components, are plotted in blue, orange, cyan, magenta, and gray, respectively}) and residuals. The spectra are extracted from the region indicated by the cyan dashed annulus in Figure \ref{fig:RGB_whole}. }\label{fig:global_bkg}}
	\end{figure*}

	We extract the global spectra of W49B from a circular region which covers the whole SNR (as indicated by the white circle in Figure \ref{fig:RGB_whole}). The spectra are shown in Figure \ref{fig:global_bkg}, and are dominated by emission lines from the He-like and H-like ions of Si, S, Ar and Ca, and by an intense Fe K line complex. \citet{2013ApJ...764...50L} reported a dramatically decline of the emission line fluxes for W49B during an 11 {yr} interval between two of the {\it Chandra} {ACIS} observations ($\sim 18\%$ for Si XIII and $\lesssim 5\%$ for other emission lines, from 2000 to 2011). However, it is difficult to figure out a proper astrophysical explanation for this line flux decrement. The {\it XMM-Newton} spectra of W49B  extracted from 2004 and 2014 observations are well consistent with each other, and show no signs of line flux decrement {(see Figure \ref{fig:global_bkg})}. We infer that a part of the flux decline indicated by {\it Chandra} observations may be caused by the quantum efficiency (QE) contamination of ACIS\footnote{The effective ACIS QE is lower than it was at launch, and is continuously decreasing. {As a reference, the effective area of ACIS-S at $\sim1.8$\,keV (Si XIII band) was $\sim640$\,cm$^2$ in 2000, while it declined to $\sim610$\,cm$^2$ in 2011 (decreased by $\sim5\%$).} This QE contamination takes place mainly in low-energy band, which may be responsible for a larger decrement of Si XIII flux than that of other emission lines. {However, a decrement of $\sim18\%$ in line flux is still confusing. Further diagnosis of the {\it Chandra} data with the newest calibration is needed, which is, however, beyond the purpose of this work.} For a comprehensive analysis of ACIS QE contamination, please refer to http://cxc.harvard.edu/ciao/why/acisqecontamN0010.html and the references therein.}. On the other hand, we find that there are some slight differences between the MOS spectra and the pn spectra (mainly in the $\lesssim 1.0$\,keV and $\gtrsim 7.0$\,keV energy bands). This may be caused by the different instrumental backgrounds of the two cameras \citep[see, e.g.,][]{2004A&A...414..767K,2008A&A...478..575K}. In view of a simpler background and a better spectral resolution of the MOS camera, we use only the MOS spectra for our further spectral analysis. 
	
	For background analysis, we extract the spectra from an annulus region around the SNR (indicated by the dashed cyan circles in Figure \ref{fig:RGB_whole}). We jointly fit the 0.3--10.0\,keV spectra using a background model which contains both the astrophysical and instrumental components. {Following} \citet{2009PASJ...61S.115M} and \citet{2013PASJ...65...19U}, the astrophysical part of the background model consists of three components: {a {\tt powerlaw} model for} the cosmic X-ray background (CXB), {and two {\tt apec} thin thermal plasma models for} the foreground emission from the local hot bubble (LHB) and the Galactic emission (diffuse X-ray emission from the Galactic halo and the Galactic ridge), {respectively}. {The CXB and the Galactic emission are subject to  absorptions which are modeled with {\tt phabs} code.}  The instrumental background mainly consists of two fluorescence lines: Al K$\alpha\sim1.49$\,keV and Si K$\alpha\sim1.74$\,keV \citep{2008A&A...478..575K}. {Another} {\tt powerlaw} component is included to account for the remaining SP contamination. All the parameters of the astrophysical components are linked between different observations, while the fluxes of the instrumental lines and the SP contamination are not, since they vary with detectors and observations. The model described above can basically reproduce the background spectra, but {leaves} a line-like residual at $\sim6.7$\,keV and a narrow bump-like residual at $\sim0.9$\,keV. The former feature is likely the Fe K emission from the high-temperature plasma in the Galactic ridge \citep{2013PASJ...65...19U}. But the physical origin of the narrow bump-like structure is unclear. Similar feature has been noticed by \citet{2009PASJ...61S.115M} based on a {\it Suzaku} observation of the Galactic plane, and they attributed it to the emission from unresolved dM stars. {To keep the background model simple, we just add gaussian components for these two features, and leave out further discussions on the detailed physical origins.}
	Therefore, the background model can be finally described as:
	\begin{equation}
	{\tt apec}_{\rm LHB} + {\tt phabs1}\times {\tt apec}_{\rm Gal} + {\tt phabs2} \times {\tt powerlaw}_{\rm CXB} + {\tt powerlaw}_{\rm SP} + \sum_{i}{\tt gauss}_i,
	\end{equation}
	{where $i=1, 2, 3, 4$ denote the two instrumental lines (Al K$\alpha\sim1.49$\,keV and Si K$\alpha\sim1.74$\,keV) and the two line-like features at $\sim0.9$\,keV and $\sim6.7$\,keV, respectively.}
	The best-fit parameters are summarized in Table \ref{tab:spec_bkg}, and the fitted spectra are shown in Figure \ref{fig:global_bkg}. The best-fit background model obtained here is then scaled by the area factors and used in {the} subsequent analysis of the source spectra.
	
	Previous studies have shown that the spectrum of W49B can be well reproduced by a two-component model which contains a low-temperature plasma with solar abundance and a high-temperature plasma with super-solar abundance, representing the shocked interstellar medium (ISM) and ejecta, respectively \citep[e.g.,][]{2000ApJ...532..970H,2006A&A...453..567M,2013ApJ...777..145L,2018AA...615A.150Z}. Moreover, the {shocked} ISM is suggested to be under the CIE, while the {shocked} ejecta is found to have some overionization features, such as the RRC of Fe and high {H-like to He-like} line ratios of S, Ar and Fe \citep[e.g.,][]{2009ApJ...706L..71O,2010AA...514L...2M,2013ApJ...777..145L,2018ApJ...868L..35Y}. 
	
\begin{figure*}
		\centering
		\includegraphics[width=0.65\textwidth]{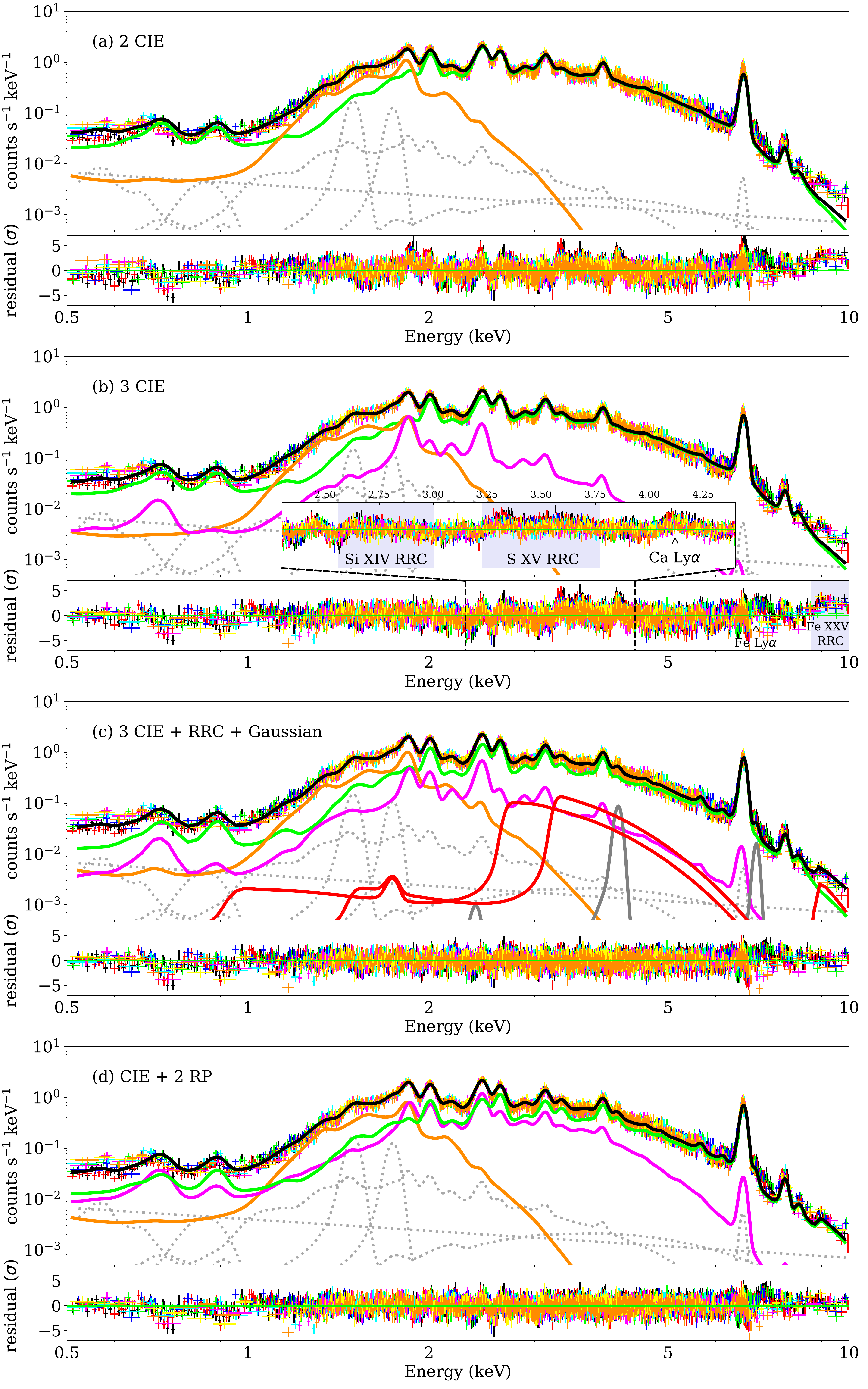}
		\caption{The global spectra of W49B (data points), with various spectral fit models (solid curves, different colors for different components) and residuals. Background components are plotted with light gray dotted curves. (a) ``2 CIE'' model, in which the orange curve denotes the ISM component and the green curve denotes the ejecta component; (b) ``3 CIE'' model, in which the magenta curve denotes the additional low-temperature ejecta component; (c) ``3 CIE + RRC + Gaussian'' model, in which RRCs are plotted in red and Ly$\alpha$ lines are plotted in dark gray; (d) ``CIE + 2 RP'' model, in which the high- and low-temperature RP components are plotted in green and magenta, respectively.\label{fig:spec_fit}}
	\end{figure*}

	We start our {spectral} analysis by fitting the 0.5--10.0\,keV global spectra of W49B with a simple two-temperature CIE model, and seeing whether additional components are needed.\footnote{\citet{2018AA...615A.150Z} reported the detection of several point-like sources in the W49B based on {\it Chandra} observations. The overall flux of them is $\lesssim10^{-5}$\,photons\,s$^{-1}$\,cm$^{-2}$ (0.7--5.0\,keV), which is about three orders of magnitude lower than the flux of W49B. Thus the effects of the unresolved point-like sources can be ignored in our spectral fitting.} The model, {as shown in Figure \ref{fig:spec_fit}a},  contains two absorbed {\tt vvapec} components: one with solar abundances accounts for the ISM-dominated plasma,  {and} the other with enhanced abundances accounts for the ejecta-dominated plasma. We find that this ``2 CIE'' model can not properly reproduce the spectra and gives a large \RCS$\sim1.903$\,(16069/8446). 
	The most significant residual turns up around He-like Si and S lines and H-like Ar and Ca lines, which indicates another temperature components may be needed. We then try a ``3 CIE'' model by adding a third {\tt vvapec} component for the ejecta-dominated plasma (Figure \ref{fig:spec_fit}b). The fitting {gets} some improvements and \RCS is reduced to $\sim1.775$\,(14987/8443). However, much of the residual in {energies} $\gtrsim2.0$\,keV still exist. Adding further plasma component can not improve the fitting anymore. By carefully checking the residual, we find some signs for overionization, including: (1) bump-like features at $\sim3.2$\,keV and $\sim8.8$\,keV, which may corresponding to the RRCs of He-like S and He-like Fe, respectively (as well as a less significant feature at $\sim2.7$\,keV which may related to the RRC of H-like Si); (2) {emission-line-like} features at $\sim4.1$\,keV and $\sim7.0$\,keV, which {appear to correspond} to Ly$\alpha$ lines of Ca and Fe. To further confirm these features, we add three {\tt redge} components (for RRCs) and two {\tt gauss} components (for Ly$\alpha$ lines) to the model (Figure \ref{fig:spec_fit}c). It gives acceptable fitting results and further reduces the \RCS to $\sim1.393$\,(11750/8436). The fluxes of individual components are obtained as (in units of $10^{-4}$\,\fluxunit): $6.70^{+0.33}_{-0.17}$ for Si XIV RRC, $5.50^{+0.13}_{-0.16}$ for S XV RRC, $1.08^{+0.20}_{-0.16}$ for Fe XXV RRC, $0.51\pm0.04$ for Ca Ly$\alpha$, and $0.22\pm0.02$ for Fe Ly$\alpha$. Moreover, we get a temperature $0.64\pm0.01$\,keV for {the} Si and S RRC, but a much higher temperature $1.08^{+0.20}_{-0.16}$\,keV for {the} Fe RRC. It should be noted that the spectrum model here is just quasi-physical (CIE + overionization features), the fluxes and temperatures obtained above may differ from the reality. However, it still provides strong evidences of the overionization state not only for the Fe-rich ejecta, but also for the Si-, S- and Ca-rich ejecta. Furthermore, it indicates a multi-temperature composition of the RP.
	
	{In view of} the analysis above, we construct a ``CIE + 2 RP'' model to finally fit the global spectra, in which the RP components are described by two {\tt vvrnei} models embedded in {XSPEC}. {The {\tt vvrnei} model characterizes the spectrum of a NEI plasma which is assumed to have started in collisional equilibrium with an initial temperature $kT_{\rm init}$ and is rapidly heated or cooled to a temperature $\kTe$.} This {``CIE + 2 RP''} model well reproduces the spectra with a \RCS$\sim1.396$\,(11784/8439). The detailed parameter setting and the best-fit results are summarized in Table \ref{tab:spec_whole}, and the fitted spectra are shown in Figure \ref{fig:spec_fit}d. Based on the results, the ejecta-dominated plasma in W49B has two major components with different thermal and ionization states. Both of the two components {are} overionized, and the current and initial electron temperatures are $kT_{\rm e,1} = 1.60^{+0.02}_{-0.01}$\,keV, $kT_{\rm e,2} = 0.64\pm0.01$\,keV and $kT_{\rm init,1} = 4.54^{+0.17}_{-0.07}$\,keV,  $kT_{\rm init,2} = 2.42^{+0.05}_{-0.03}$\,keV, respectively. The ionization parameters are $n_{\rm e,1}t_1 = 3.90^{+0.08}_{-0.04}\times10^{11}$\,cm$^{-3}$\,s and $n_{\rm e,2}t_2 = 5.49^{+0.04}_{-0.09}\times10^{11}$\,cm$^{-3}$\,s.
	
	\subsection{EW map}\label{sec:EW}
	
	EW map is a powerful tool to investigate the spatial distribution of the {true} line strength while avoiding the contamination of the underlying continuum \citep[e.g.,][]{2000ApJ...537L.119H}. \citet{2006A&A...453..567M} produced EW maps of S, Ar, Ca and Fe for W49B based on the $\sim30$\,ks {XMM-Newton} data ({i.e.} the first two observations listed in Table \ref{tab:obs}). However, their results were limited by the exposure time and total net counts. The additional $\sim150$\,ks observations can help us carry out EW analysis with much more details. 
	
	In order to define the energy ranges for individual emission lines, we carefully analyze the 1.5--7.0\,keV spectra by fitting them with a phenomenological model which consists of two thermal bremsstrahlung continuum and several Gaussian lines. We obtain the energy ranges for all the He$\alpha$ and Ly$\alpha$ lines of Si, S, Ar and Ca, for He$\alpha$ lines of Cr and Mn, and for Fe K complex ({as} summarized in Table \ref{tab:range}). For each emission line, we select a low-energy continuum $C_{\rm low}$ and a high-energy continuum $C_{\rm high}$, and use a linear interpolation (or extrapolation) between them to estimate the underlying continuum. The {true} line flux is obtained by subtracting the underlying continuum. Then, the EW can be calculated as the ratio between the truly line flux and the continuum flux density at the line centroid. It should be noted that the continuum-subtracted line flux represents the emission measure of a certain ion, which is proportional not only to the ion density but also to the electron density and the emission volume. On the other hand, the EW is {linearly} related to the abundance but is still affected by the temperature and the ionization state.
	
\begin{figure*}
		\plotone{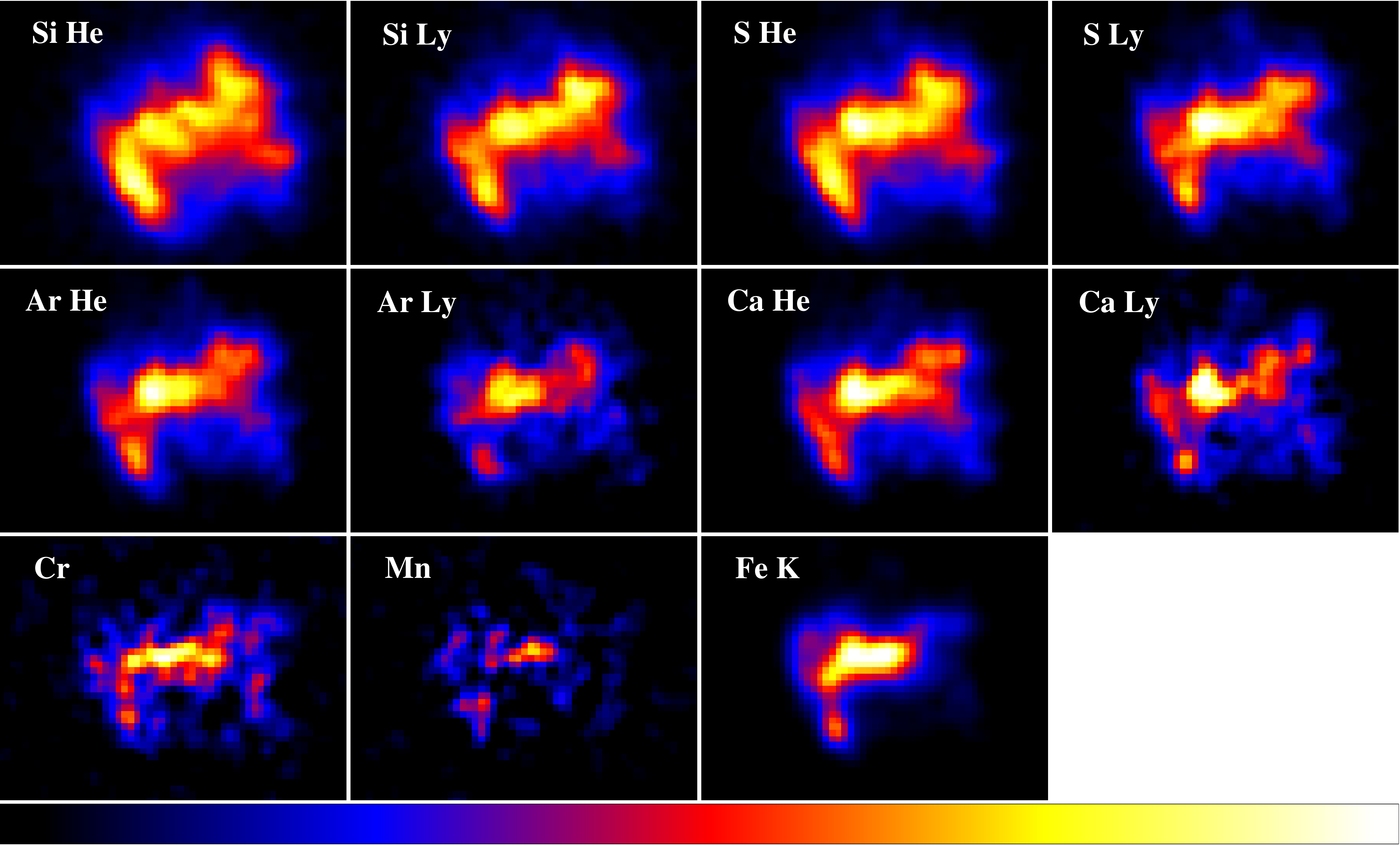}
		\caption{Vignetting-corrected, QPB- and continuum-subtracted images of line fluxes. The flux range between 0 and: 70 (Si He); 60 (Si Ly); 100 (S He); 70 (S Ly); 70 (Ar He); 30 (Ar Ly); 60 (Ca He); 20 (Ca Ly); 6 (Cr); 5 (Mn); 180 (Fe K), in units of \LFunit. The color bar has a linear scale.\label{fig:line_flux}}
	\end{figure*}

\begin{figure*}
		\plotone{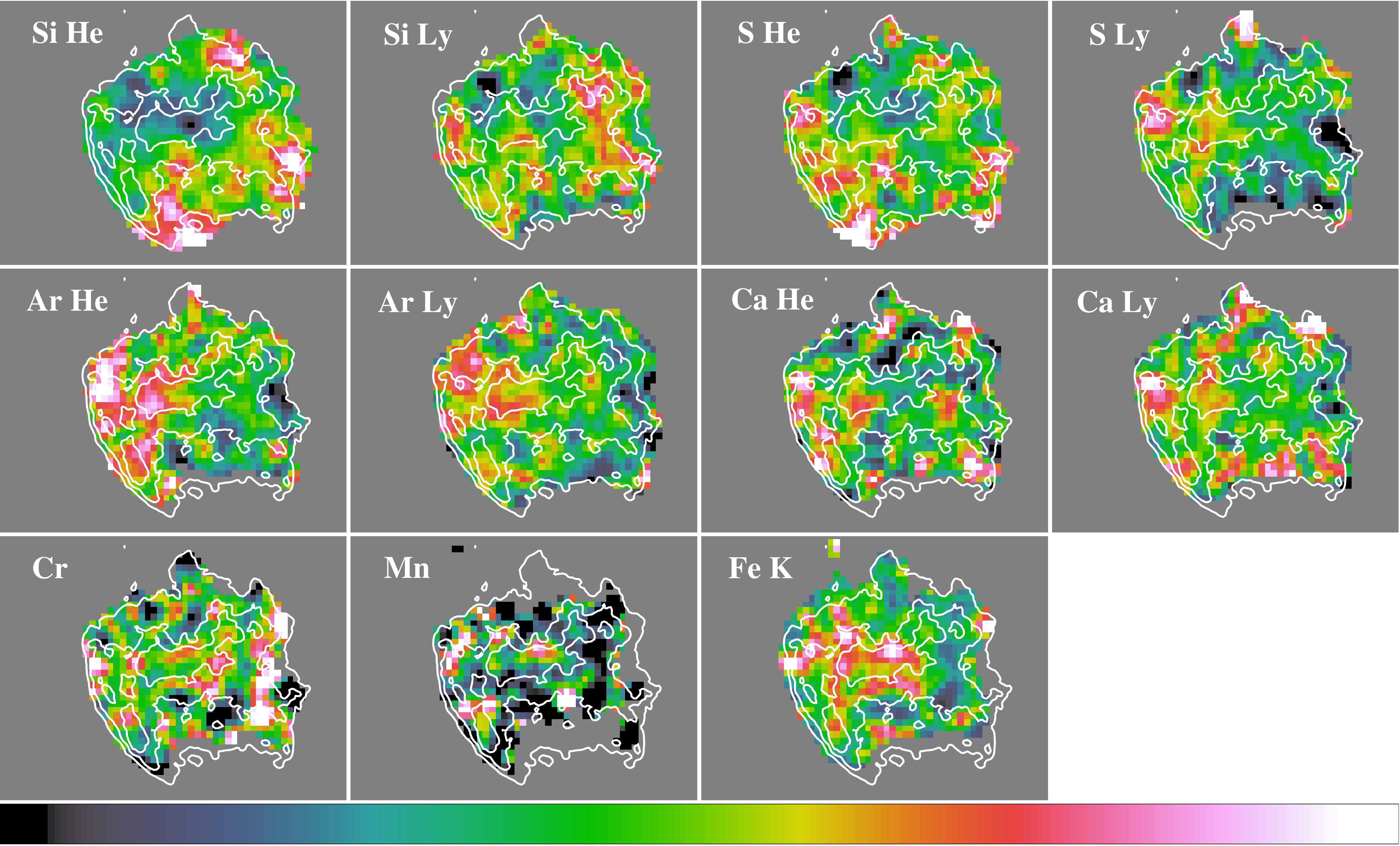}
		\caption{EW maps of emission lines. The EW range between: 100--300 (Si He); 80--200 (Si Ly); 150--350 (S He); 80--250 (S Ly); 80--200 (Ar He); 0--100 (Ar Ly); 120--240 (Ca He); 0--120 (Ca Ly); 0--120 (Cr); 0--120 (Mn); 0--8500 (Fe K), in units of eV. The color bar has a linear scale. The white contours denote the {\it Chandra} 0.5--8.0\,keV flux \citep{2013ApJ...764...50L}.\label{fig:EW_map}}
	\end{figure*}

	The vignetting-corrected, QPB- and continuum-subtracted images of line fluxes are shown in Figure \ref{fig:line_flux}. The EW maps are shown in Figure \ref{fig:EW_map}. The images are produced using the SAS task {\tt adapt\_merge}, which rebins the images to a pixel size of $0'.12\times0'.12$ and adaptively smooths the images with a minimum count of 25. Similar to \citet{2006A&A...453..567M}, we find a bright central bar-like structure in all the line flux images, which is also the most luminous part of W49B in {X-rays}. The line fluxes of intermediate mass elements (IMEs) such as Si, S, Ar and Ca show similar distributions with bright knots present at the both sides of the central bar. Fe K complex has a peculiar distribution concentrated in the northeast part of the remnant while is almost unseen in the southwest. Mn shows a similar distribution {to} Fe K. On the other hand, EW maps reveal unique properties which are quite different from the line fluxes. A central-bright pattern is only preserved {in} the EW {maps} of Fe K and Mn, while for the IMEs, high EWs appears mainly on the periphery of the remnant. This indicates that although the central bar-like structure has the highest emission measure for almost all the emission lines, the {true} distributions of the metal abundances show stratified features. 
	
	
	\section{Discussion} \label{sec:disc}
	
	\subsection{Density and mass of the X-ray emitting plasma}
	Based on the normalization parameters\footnote{Defined as $10^{-14}/(4\pi d^2)\int n_{\rm e} n_{\rm H} f dV$, where $d$ is the distance to the remnant, $n_{\rm e}$ and $n_{\rm H}$ are the electron and hydrogen density, $f$ is the filling factor; $n_{\rm e}=1.2n_{\rm H}$ for a fully ionized plasma with solar abundance.} obtained in global spectral fitting, we further estimate the electron densities of individual plasma component in W49B. Assuming a spherical geometry of the remnant with a radius of $\sim2'.2$, the electron densities are obtained as:  $n_{\rm e,1}=0.46\pm0.01f_1^{-0.5}d_{9.3}^{-0.5}$\,cm$^{-3}$, $n_{\rm e,2}=0.62\pm0.01f_2^{-0.5}d_{9.3}^{-0.5}$\,cm$^{-3}$ and $n_{\rm e,3}=18.17\pm0.06f_3^{-0.5}d_{9.3}^{-0.5}$\,cm$^{-3}$ (hereafter, we use the subscripts 1, 2 and 3 to indicate the parameters related to the high-temperature RP, the low-temperature RP and the CIE component, respectively), where $d_{9.3}=d/(9.3\,{\rm kpc})$ is the distance scaled to 9.3\,kpc. If we further assume that the three plasma components are in pressure balance (i.e., $n_{\rm e,1}kT_{\rm e,1}=n_{\rm e,2}kT_{\rm e,2}=n_{\rm e,3}kT_{\rm e,3}$) and sharing the whole volume (i.e., $f_1+f_2+f_3=1$), the densities and filling factors can be calculated as: $n_{\rm e,1}=2.1\pm0.1d_{9.3}^{-0.5}$\,cm$^{-3}$, $n_{\rm e,2}=5.2\pm0.3d_{9.3}^{-0.5}$\,cm$^{-3}$, $n_{\rm e,3}=18.8\pm0.1d_{9.3}^{-0.5}$\,cm$^{-3}$ and $f_1\sim5.0\%$, $f_2\sim1.4\%$, $f_3\sim93.6\%$, respectively. The parameters obtained above indicate that the X-ray emitting volume in W49B is dominated by {the shocked} ISM, while the metal-rich RP components distribute in a rather small part of the remnant (with a total filling factor $<10\%$). The total masses of the X-ray emitting gas are $M_1=2.7\pm0.4d_{9.3}^{2.5}\,\Msun$, $M_2=1.9\pm0.3d_{9.3}^{2.5}\,\Msun$ and $M_3=452\pm5d_{9.3}^{2.5}\,\Msun$. 
	
	\citet{2005ApJ...631..935K} constrained the masses of the high- and low-temperature plasma in W49B to be $M_{\rm HT}\sim34d_{9.3}^{2.5}\Msun$ and $M_{\rm LT}\sim138d_{9.3}^{2.5}\Msun$, {respectively}, based on {\it Chandra} observations. Recently, \cite{2018AA...615A.150Z} carried out a spatially resolved spectroscopic study of W49B, and obtained the gas masses as $M_{\rm HT}\sim43d_{9.3}^{2.5}\Msun$ and $M_{\rm LT}\sim404d_{9.3}^{2.5}\Msun$. We note that the above works suggested a much higher mass for the high-temperature, ejecta-dominated plasma than {that} we obtained here ($M_1+M_2\sim4.6d_{9.3}^{2.5}\Msun$). 
	We consider this as a result of the different {spectral} models adopted in different works. As {analyzed} in Section \ref{sec:spec_fit}, the global spectra of W49B are best characterized by a three-temperature plasma model ($kT_{\rm e,1}\sim1.60$\,keV, $kT_{\rm e,2}\sim0.64$\,keV, $kT_{\rm e,3}\sim0.18$\,keV), while both of the above works used a two-temperature model to fit the spectra ($kT_{\rm e,HT}\sim0.6$--$2.2$\,keV, $kT_{\rm e,LT}\sim0.27$\,keV). As a result, the electron temperature of the cold ISM-dominated component may be overestimated, while the normalization be underestimated in {these} works. Our analysis reveals a rather large normalization parameter for the low-temperature ISM, which is more than two orders of magnitude higher than that of the shocked ejecta. This then naturally results in a rather small filling factor and gas mass of the ejecta-dominated plasma according to the calculations above.
	\subsection{Ionization temperature}
\begin{figure*}
		\fig{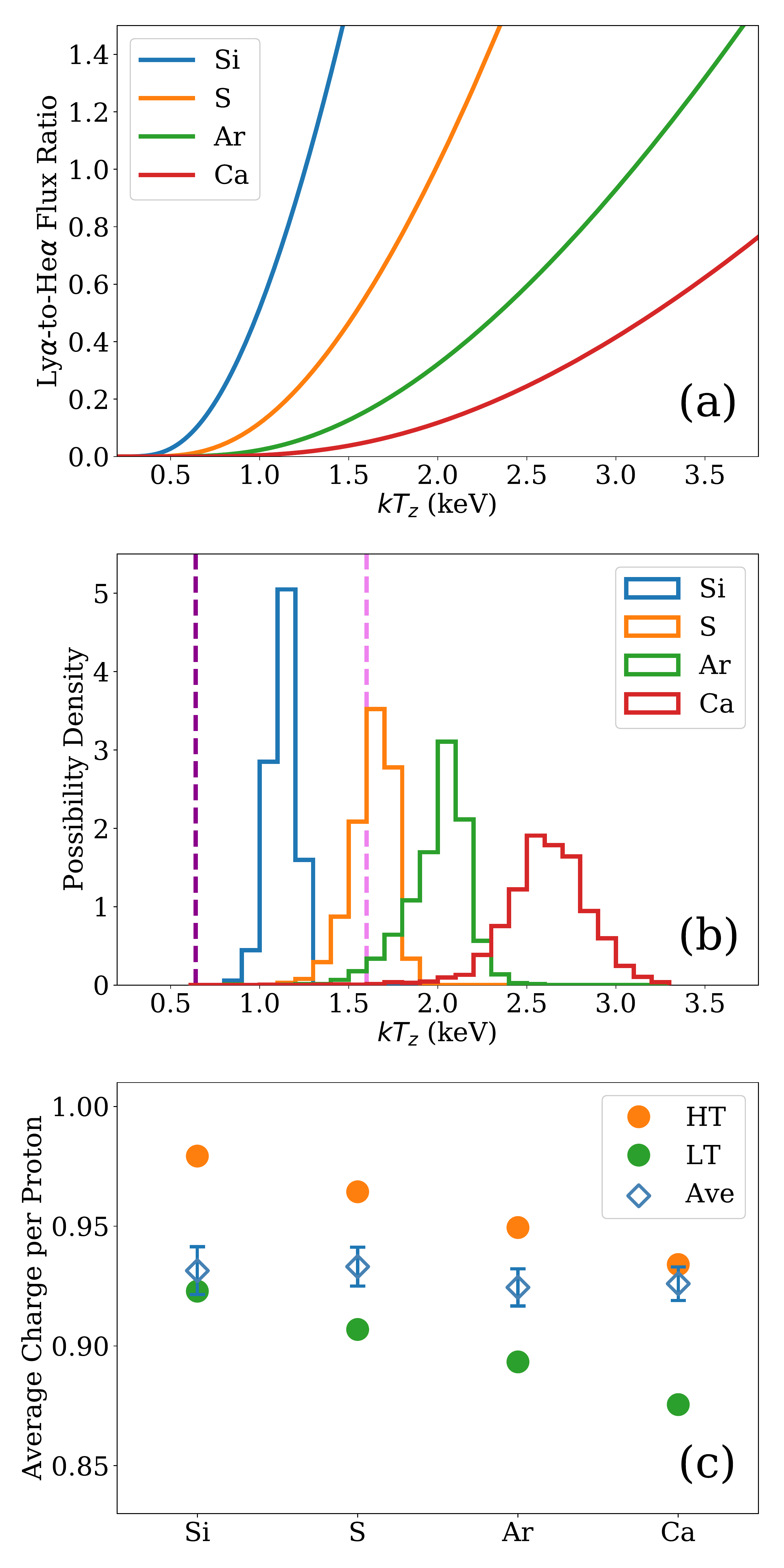}{0.5\textwidth}{}
		\caption{(a) The modeled relations between the Ly$\alpha$-to-He$\alpha$ flux ratios and $\kTz$; (b) the flux-weighted possibility distribution functions of the $\kTz$ for Si, S, Ar and Ca, {the dark and light magenta dashed line indicate the $\kTe$ of two RP components ($\sim$0.64\,keV and $\sim$1.6\,keV), respectively}; (c) the average charges of Si, S, Ar and Ca calculated from the flux ratios, {compared with those calculated from the best-fit parameters of the high-temperature (HT) and the low-temperature (LT) RP} components obtained in Section \ref{sec:spec_fit}. \label{fig:kTz_plot}}
	\end{figure*}

\begin{figure*}
		\plotone{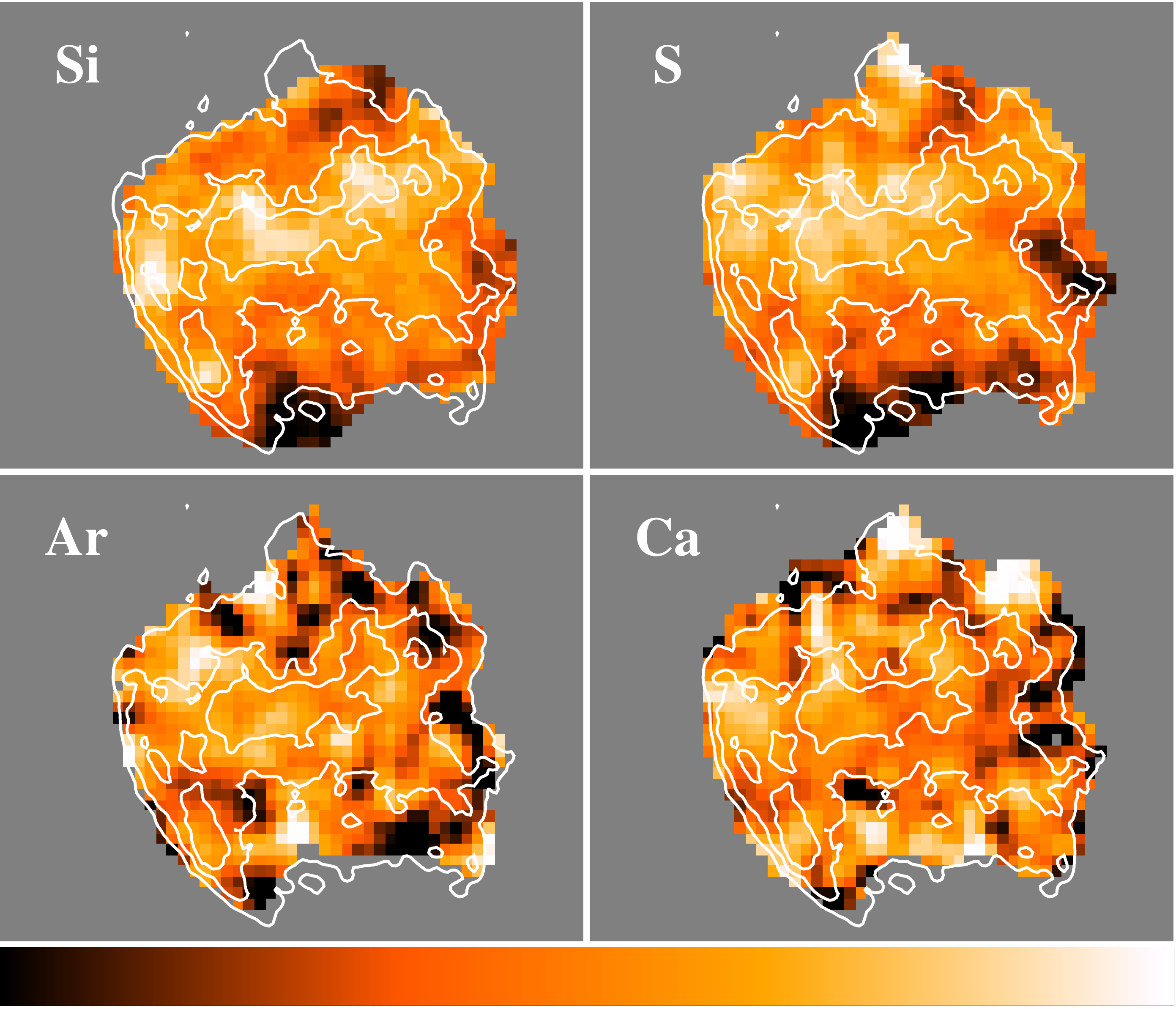}
		\caption{Distribution maps of $\kTz$ for Si, S, Ar and Ca in W49B. $\kTz$ range between 0.9--1.3\,keV for Si, 1.3--1.9\,keV for S, 1.6--2.4\,keV for Ar, and 2.0--3.2\,keV for Ca. The color bar has a linear scale. \label{fig:kTz_map}}
	\end{figure*}

	The ionization temperature ($\kTz$) is commonly used to describe the ionization state of the plasma. It can be estimated by the flux ratios between H-like and He-like emission lines \citep[e.g.,][]{2002ApJ...572..897K,2009ApJ...706L..71O,2013ApJ...777..145L}, {but may vary from one ion species to another}. Here, we estimate the ionization temperatures for Si, S, Ar and Ca based on their Ly$\alpha$ and He$\alpha$ line flux obtained in Section \ref{sec:EW}. The modeled relations between the Ly$\alpha$-to-He$\alpha$ flux ratios and $\kTz$ are shown in Figure \ref{fig:kTz_plot}a, which are derived from the SPEX code\footnote{https://www.sron.nl/astrophysics-spex} \citep{1996uxsa.conf..411K}. 
	With these relations, we estimate the ionization temperatures and their distributions in W49B, as shown in Figure \ref{fig:kTz_map}.
	
	We further calculate the flux-weighted possibility distribution functions of the $\kTz$ for different ion species based on their line flux and $\kTz$ maps, as shown in Figure \ref{fig:kTz_plot}(b). We find that the average ionization temperatures of different ion species are indeed different: $\overline{kT}_{\rm z,Si}=1.13\pm0.08$ for Si, $\overline{kT}_{\rm z,S}=1.63\pm0.12$ for S, $\overline{kT}_{\rm z,Ar}=2.00\pm0.17$ for Ar, and $\overline{kT}_{\rm z,Ca}=2.62\pm0.25$ for Ca ({here, we give the flux-weighted averages and their standard deviations}). Additionally, the overionization states of Si, S, Ar and Ca can be further evidenced by comparing their average ionization temperatures with the electron temperatures obtained before ($kT_{\rm e,1}\approx1.60$\,keV and $kT_{\rm e,2}\approx0.64$\,keV). \citet{2013ApJ...777..145L} analyzed the Chandra spectra of 13 regions in W49B, and obtained the ionization temperatures of S and Ar as $\overline{kT}_{\rm z,S}\approx1.3$--$1.7$\,keV and $\overline{kT}_{\rm z,Ar}\approx1.5$--$2.1$\,keV, {which are consistent with the results of this work.}
	
	The difference of $\kTz$ among ion species may result from the different initial conditions and recombination history \citep[e.g.,][]{2012PASJ...64...81S}. In the case of W49B, we find that the $\kTz$ of Si and S are lower than those of Ar and Ca. This seems to be unphysical for a single temperature RP, given that the recombination timescales of Si and S are longer than those of Ar and Ca under a temperature of $\sim10^7$\,K \citep{2010ApJ...718..583S}, as pointed out by \citet{2013ApJ...777..145L}. However, it has been shown that the RP in W49B consists of at least two major components with different temperatures and ionization parameters (Section \ref{sec:spec_fit}). Thus the difference of $\kTz$ can be reasonably understood as a result of the different distributions of ion populations among different RP components. Figure \ref{fig:kTz_plot}c shows the average charges of different ion species calculated with SPEX code, using the {flux-ratio-inferred} average $\kTz$ obtained above, and the best-fit parameters ({i.e. electron temperatures and ionization parameters}) of {the} two RP {components} obtained in Section \ref{sec:spec_fit}, respectively. We find that the flux-ratio-inferred and spectral-fitting-inferred average charges consist well with each other, which strengthens the reliability of both approaches. Moreover, Figure \ref{fig:kTz_plot}c indicates that Si and S are dominated in the low-temperature RP while Ar and Ca are dominated in the high-temperature RP, which may be responsible for a lower $\kTz$ of Si and S.
	
	\subsection{Recombination age and the origin of the RP}
	With the electron density $n_{\rm e}$ and the ionization parameter $n_{\rm e}t$ given, we can estimate the elapsed time since the RP was formed, which is the so-called recombination age ($t_{\rm rec}$). At a distance of 9.3\,kpc, the recombination ages are derived as $t_{\rm rec,1}=6000\pm400$\,yr for the high-temperature RP and $t_{\rm rec,2}=3400\pm200$\,yr for the low-temperature RP, respectively. {As a comparison}, \citet{2018AA...615A.150Z} obtained similar results with $t_{\rm rec}\sim2000$--$6000$\,yr, {based on {\it Chandra} observations}. 
	
	The physical origin of the RP presented in SNRs has not been clearly understood. As mentioned above, there are two major scenarios that may lead to rapid electron cooling and result in overionization: adiabatic expansion \citep[e.g.,][]{1989MNRAS.236..885I} and thermal conduction {\citep[e.g.,][]{2002ApJ...572..897K,2011MNRAS.415..244Z}}. In the case of W49B, {there is already some observational evidence}
	for a gradient of increasing overionization degree from east to west, which supports an adiabatic expansion origin of RP \citep[e.g.,][]{2010AA...514L...2M,2013ApJ...777..145L,2018ApJ...868L..35Y}. On the other hand, the thermal conduction between ambient clouds and hot plasma may also play an important role to form the RP, as revealed by \citet{2011MNRAS.415..244Z} and \citet{2019ApJ...875...81Z} based on hydrodynamic simulations. Observationally, one can investigate the possibility of the thermal conduction origin of RP by comparing the conduction timescale $t_{\rm cond}$ with the recombination age $t_{\rm rec}$ \citep[e.g.,][]{2002ApJ...572..897K,2012PASJ...64..141U,2014PASJ...66..124S}. The {classical} conduction timescale can be estimated as
	\begin{equation}
	t_{\rm cond}\approx 2\times10^{10}\left(\frac{n_{\rm e}}{1\,{\rm cm}^{-3}}\right)\left(\frac{l_{\rm T}}{1\,{\rm pc}}\right)^2\left(\frac{kT_{\rm e}}{1.0\,{\rm keV}}\right)^{-5/2}\left(\frac{\ln\Lambda}{32.2}\right)~{\rm s},
	\end{equation}
	where $l_{\rm T}\equiv({\rm grad}\,\ln T)^{-1}$ is the temperature gradient scale length and $\ln\Lambda$ is the Coulomb logarithm {\citep{1962pfig.book.....S,2002ApJ...572..897K,2014ApJ...791...87Z}}. For the high-temperature RP component with an initial temperature $kT_{\rm init,1}\sim4.5$\,keV and an electron density $n_{\rm e,1}\sim2.1$\,cm$^{-3}$, it can efficiently cool down by thermal conduction in a timescale of $t_{\rm cond,1}\sim1100\,{\rm yr}<t_{\rm rec,1}$. But for the low-temperature RP component with $kT_{\rm init,2}\sim2.4$\,keV and $n_{\rm e,2}\sim5.2$\,cm$^{-3}$, the timescale is $t_{\rm cond,2}\sim13000\,{\rm yr}>t_{\rm rec,2}$. Here, we assume $l_{\rm T}\sim 6$\,pc, which is comparable to the SNR radius.
	Therefore, the thermal conduction scenario is a possible origin for the high-temperature RP in W49B, while it may not apply to the low-temperature RP due to a much longer cooling timescale. 
	{If the thermal conduction takes place on a smaller scale, where the temperature gradient scale length $l_{\rm T}$ is comparable or even shorter than  the electron mean free path $\lambda_{\rm e}\approx 0.4\left(\kTe/1.0\,{\rm keV}\right)^{2}\left(n_{\rm e}/1\,{\rm cm}^{-3}\right)^{-1}\,{\rm pc}$, the heat flux becomes saturated \citep{1977ApJ...211..135C}. However, the hydrodynamic simulations which include the effect of saturated thermal conduction show that the conduction timescale can still be a few $\times10^3$\,yr \citep[e.g.,][]{2011MNRAS.415..244Z}, and thus makes little effect on our conclusions here.
	}
	
	Actually, the adiabatic expansion and the thermal conduction may simultaneously contribute to the formation of the RP, especially when the SNR associates with complex cloud environment, such as in the case of W49B \citep[e.g.,][]{2011MNRAS.415..244Z,2019ApJ...875...81Z}.
	
	\subsection{SN type of W49B}
	The SN type of W49B is still under debate. It is usually considered as a remnant of CC SN, based on its unique morphology and environment, the metal abundances, and the Fe K line centroid \citep[e.g.,][]{2006A&A...453..567M,2007ApJ...654..938K,2013ApJ...764...50L,2014ApJ...785L..27Y}. Moreover, the progenitor seems to be a supermassive star ($\gtrsim25\Msun$) which produced a black hole rather than a neutron star (NS) \citep{2013ApJ...764...50L}. 
	{However, the rather small cavity size \citep[radius $\sim5$--$7$\,pc, see, e.g.,][]{2007ApJ...654..938K,2014IAUS..296..170C} disfavors a supermassive progenitor. On the other hand, the CC origin has been recently doubted by \citet{2018AA...615A.150Z}, who pointed out the metal abundances are better described by SN Ia models.}
	
\begin{figure*}
		\plotone{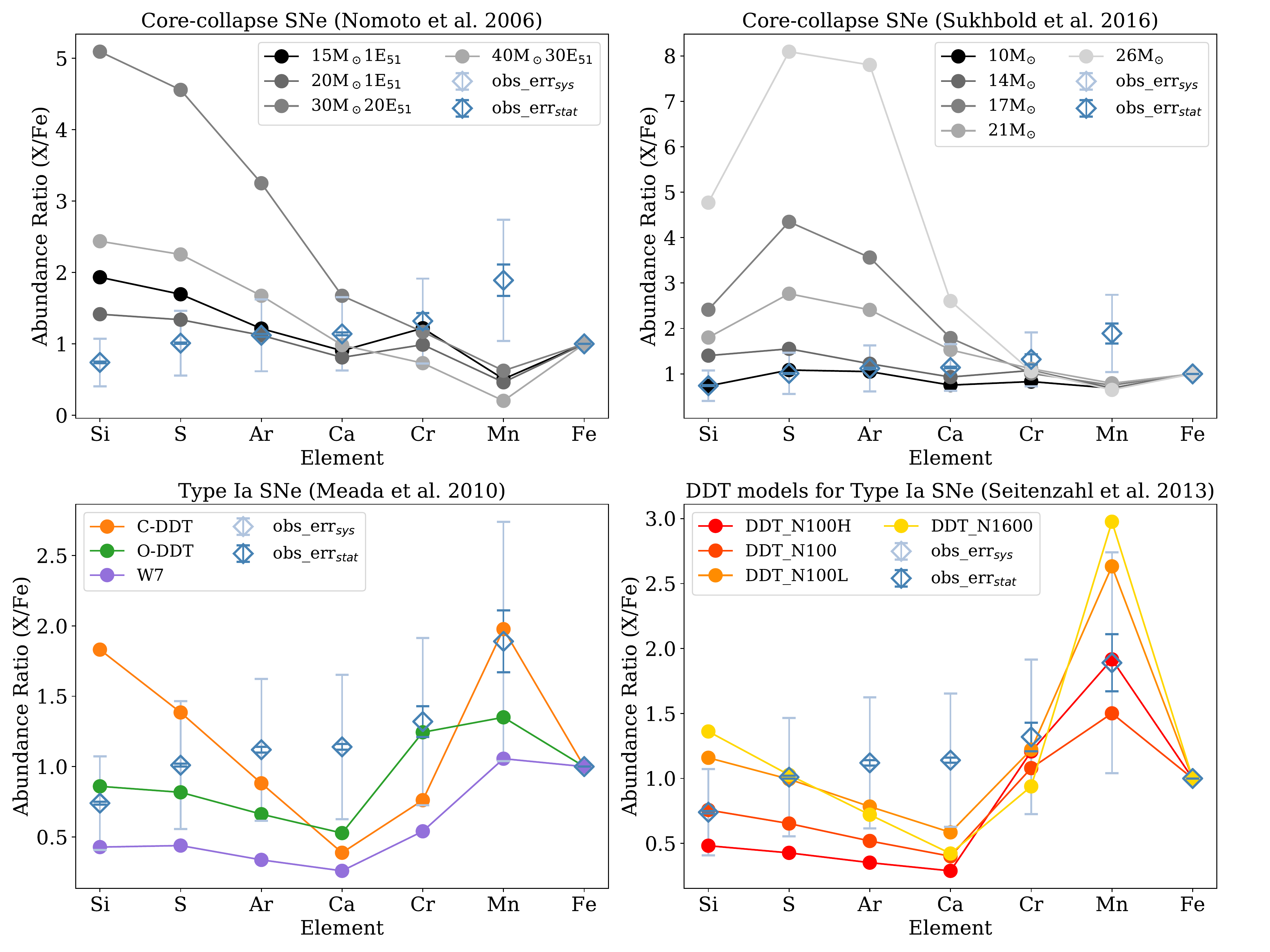}
		\caption{The metal abundance ratios of W49B ({empty diamonds, the error bars in dark blue indicate the statistical errors, while those in light blue indicate the systematic errors}), compared with the predicted results of different SN nucleosynthesis models (filled circles). Please refer to the text for a detailed description of the SN models.\label{fig:abun_ratio}}
	\end{figure*}

	In Figure \ref{fig:abun_ratio}, we compare the metal abundance ratios of the ejecta in W49B to the predicted results of different SN nucleosynthesis models. Based on the spectral fitting results, the abundance ratios {of the RP components} are obtained as: ${\rm Si/Fe}=0.74\pm0.01$, ${\rm S/Fe}=1.01\pm0.01$, ${\rm Ar/Fe}=1.12\pm0.02$, ${\rm Ca/Fe}=1.14\pm0.02$, ${\rm Cr/Fe}=1.32\pm0.11$, and ${\rm Mn/Fe}=1.89\pm0.22$. The SN models considered here include: normal CC SN and energetic hypernova models with different stellar masses and explosion energies \citep{2006NuPhA.777..424N}; typical spherical CC SN models with different stellar masses \citep{2016ApJ...821...38S}; a 1D deflagration SN Ia model W7 and 2D delayed-detonation (DDT) SN Ia models with center/off-center ignition \citep{2010ApJ...712..624M}; and 3D DDT SN Ia models with different multi-spot ignition setups \citep{2013MNRAS.429.1156S}. We find that neither CC SN models nor Type Ia SN models can perfectly reproduce the observations. The abundances of the IMEs such as Si, S, Ar and Ca favor a CC explosion, and can be well described by 10--14\,$\Msun$ progenitor models. But none of the CC SN models can reproduce the Mn-to-Fe abundance ratio as high as the observation. On the other hand, some of the Type Ia SN models such like the multi-spot ignited 3D DDT models \citep{2013MNRAS.429.1156S} may explain the high Cr and Mn abundances, while they systematically underestimate the abundances of Ar and Ca.
	
\begin{figure*}
		\plotone{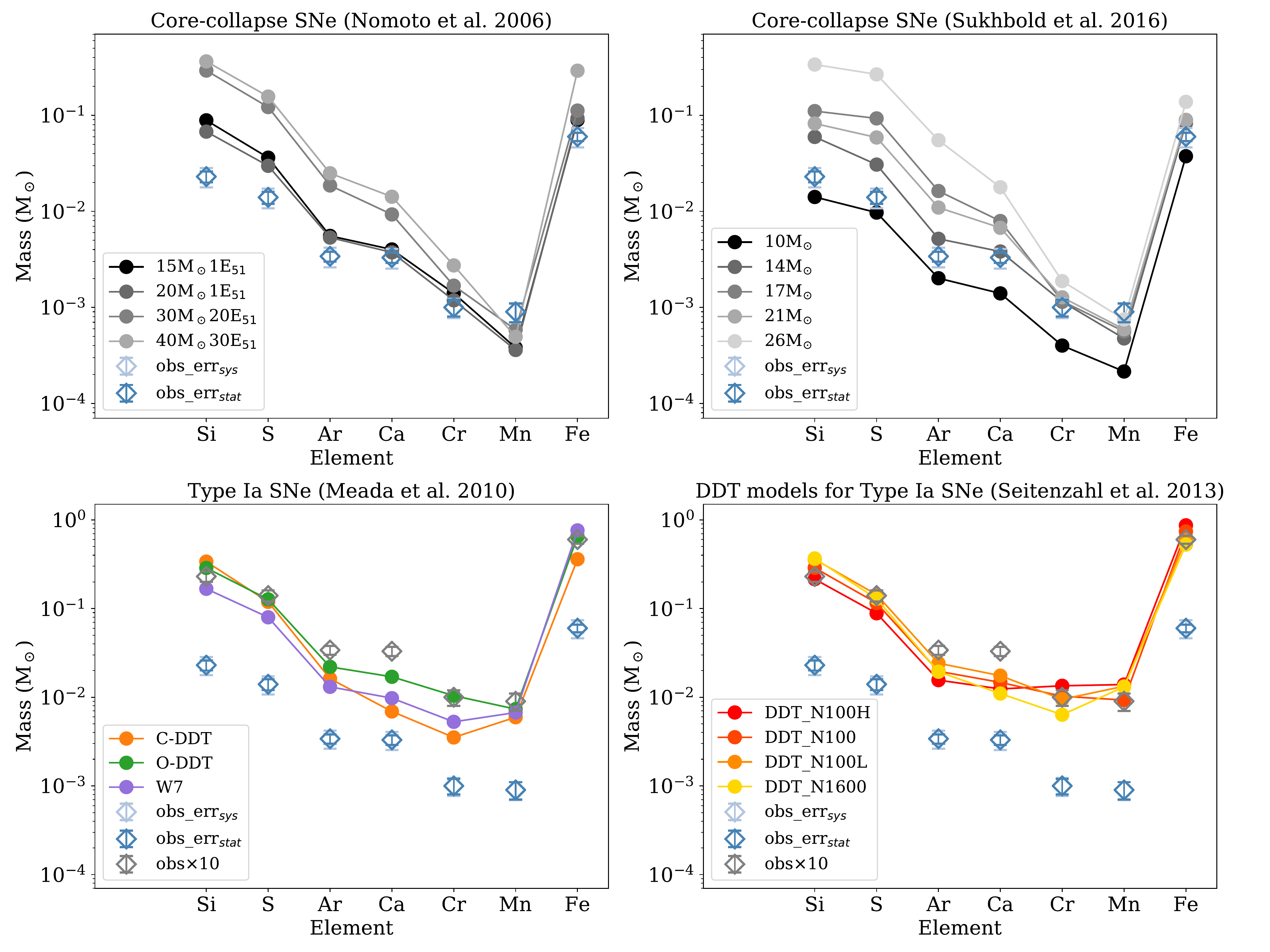}
		\caption{Similar to Figure \ref{fig:abun_ratio}, but the metal masses of the shocked ejecta in W49B. \label{fig:metal_mass}}
	\end{figure*}

	We then turn to comparing the observed masses of different metal species in ejecta with the model predicted results. The metal masses are derived from the total ejecta mass and the abundances as: $M_{\rm Si}=(2.3\pm0.3)\times10^{-2}\,\Msun$, $M_{\rm S}=(1.4\pm0.2)\times10^{-2}\,\Msun$, $M_{\rm Ar}=(3.4\pm0.4)\times10^{-3}\,\Msun$, $M_{\rm Ca}=(3.3\pm0.4)\times10^{-3}\,\Msun$, $M_{\rm Cr}=(1.0\pm0.2)\times10^{-3}\,\Msun$, $M_{\rm Mn}=(0.9\pm0.2)\times10^{-3}\,\Msun$, and $M_{\rm Fe}=(6.0\pm0.6)\times10^{-2}\,\Msun$. We note that only the X-ray emitting gas mass is counted here, which is potentially lower than the total mass produced by SN explosion. As shown in Figure \ref{fig:metal_mass}, the observed metal masses are basically consistent with the yields of CC SN models and indicate a progenitor mass of 10--14\,$\Msun$. However, the Mn mass is higher than the predicted values of all the CC SN models, with a factor of 2--5. Type Ia SN models produce much more metals than observations, for all the species considered here. If we multiply the observed masses by a factor of 10 (the gray data points in Figure \ref{fig:metal_mass}), they {would get close to the yields in} 
	the SN Ia models, except for Ar and Ca.
	
	As a conclusion, we have not {found} an SN model that can properly describe all the observational features. The progenitor of W49B {remains} unclear. However, our results provide new implications and constraints on it, which are described below.
	\begin{enumerate}
		\item If we consider W49B as a CC SNR, the analysis above indicates that the progenitor is more likely a $\lesssim15\,\Msun$ star rather than a supermassive star suggested by \citet{2013ApJ...764...50L}. A relatively small progenitor mass ($\sim13\,\Msun$) is also indicated by the small cavity size {\citep{2014IAUS..296..170C}}. In addition, nucleosynthesis in energetic explosions of the supermassive stars is characterized by smaller (Cr, Mn)/Fe than normal SNe \citep[e.g.,][]{2006NuPhA.777..424N}, which disagrees with what we have found in W49B. However, an unusual high abundance of Mn in W49B is hard to understand in a CC scenario. In CC SNe, Mn is mainly produced in incomplete explosive Si-burning conditions. Mn yield and the final Mn/Fe ratio increase with the increasing initial metallicity of the progenitor, and may be affected by detailed explosion mechanism \citep[see, e.g.,][for recent reviews on SN nucleosynthesis]{2013ARA&A..51..457N,2018SSRv..214...62T}. However it seems unlikely to have ${\rm Mn/Fe}>1$ according to existing CC SN models. Another problem for the CC scenario is the undetected stellar remnant. The SN explosion of a $\lesssim15\,\Msun$ star will leave a newborn NS which radiates X-ray emission, while no detection of such source has been confirmed yet. Recently, \citet{2018AA...615A.150Z} reported the detection of 24 point-like {X-ray} sources in the vicinity of W49B and found that nine of them are in the luminosity range predicted for an NS at an age of 5--6\,kyr according to the modeled cooling curve. Therefore, they suggested there is still the possibility that W49B harbors a cooling NS.
		\item If we consider W49B as an Ia SNR, the metal abundances we obtained above favor a DDT explosion with multi-spot ignition. The relatively small metal masses indicate that the X-ray emitting ejecta in W49B only account for $\sim10\%$ of the total SN ejecta. The rest of the ejecta may remains unshocked, or not be hot enough to emit X-ray. 
		{\citet{2014ApJ...793...95Z} found two dust components associated with W49B based on infrared observations, including a hot component $\sim 150$\,K and a warm component $\sim45$\,K. The masses of the two dust components are estimated as $7.5\pm6.6\times10^{-4}\Msun$ and $6.4\pm3.2\Msun$, respectively. However, the origins of the dust are suggested to be the swept up circumstellar or interstellar materials and the evaporation of the clouds interacting with W49B, rather than the SN ejecta \citep{2014ApJ...793...95Z}.  }
	\end{enumerate}

	{
	We note that the statistical errors of the metal abundances obtained by spectral fitting are very small (see Table \ref{tab:spec_whole}). In this case, the systematic errors may be more important, which have not been considered above. The systematic errors of the fitting results may origin from multiple aspects, such as the plasma code, the atomic database, and the fitting technique, which make it very difficult to tightly constrain the errors \citep[see, e.g., ][for a detailed discussion]{2018PASJ...70...12H}. A recent study carried out by \citet{2019arXiv191109684M} compared the SPEXACT v3.0.5 ({\tt cie} model) with the AtomDB v3.0.9 ({\tt vapec} model) and indicated systematic differences up to $\sim20\%$ for the Fe abundance and up to $\sim45\%$ for the O/Fe, Mg/Fe, Si/Fe, and S/Fe ratios. Therefore, we simply adopt a 20\% systematic error for all the metal abundances and a 45\% systematic error for all the abundance ratios considered in this work (see the light blue error bars in Figure \ref{fig:abun_ratio} and Figure \ref{fig:metal_mass}), and find that it can barely alter the conclusions obtained above.
	}

	
	\section{Summary} \label{sec:sum}
	{
	We perform a comprehensive X-ray spectroscopy and imaging analysis of SNR W49B using archival {\it XMM-Newton} data. We find {spectral} evidences of overionization {in the ejecta-dominated hot plasma} not only for Fe, but also for the lighter elements such as Si, S, and Ca. The thermal and ionization properties of the RP are well constrained, which provide new implications for the RP origin and the SN progenitor. The detailed results are summarized as follows.
	\begin{enumerate}
		\item The overionization state of the shocked-ejecta in W49B is clearly indicated by the spectral features of (1) RRCs of H-like Si, He-like S, and He-like Fe; (2) Ly$\alpha$ lines of Ca and Fe. We find that the RP has a multi-temperature composition, and the global spectra {of the remnant} can be well reproduced by a ``CIE + 2 RP'' model. The two RP components are characterized by different electron temperatures and ionization parameters as $kT_{\rm e,1} = 1.60^{+0.02}_{-0.01}$\,keV, $n_{\rm e,1}t_1 = 3.90^{+0.08}_{-0.04}\times10^{11}$\,cm$^{-3}$\,s; and $kT_{\rm e,2} = 0.64\pm0.01$\,keV, $n_{\rm e,2}t_2 = 5.49^{+0.04}_{-0.09}\times10^{11}$\,cm$^{-3}$\,s, respectively.
		\item We construct the line flux images and the EW maps of various emission lines for W49B. The results indicate that the central bar-like structure has the highest emission measure for almost all the emission lines, while the distributions of metal abundances show clear stratified features: the IMEs like Si and S distribute mainly in the outer regions, while the IGEs like Fe and Mn concentrate mainly in the inner part.
		\item The X-ray emitting volume in W49B is dominated by the shocked ISM, while the ejecta-dominated RP components distribute in a rather small part of the remnant (with a total filling factor $<10\%$). The total masses of the X-ray emitting gas are $2.7\pm0.4d_{9.3}^{2.5}\,\Msun$, $1.9\pm0.3d_{9.3}^{2.5}\,\Msun$ and $452\pm5d_{9.3}^{2.5}\,\Msun$ {for the two RP components and the shocked ISM, respectively}. 
		\item Based on the Ly$\alpha$-to-He$\alpha$ flux ratios, we estimate the average ionization temperatures of Si, S, Ar, and Ca as:  $\overline{kT}_{\rm z,Si}=1.13\pm0.08$, $\overline{kT}_{\rm z,S}=1.63\pm0.12$, $\overline{kT}_{\rm z,Ar}=2.00\pm0.17$, and $\overline{kT}_{\rm z,Ca}=2.62\pm0.25$. Combined with the electron temperatures obtained above ( $kT_{\rm e,1} \sim 1.60$\,keV and $kT_{\rm e,2} \sim 0.64$\,keV), the ionization temperatures provide further evidences of the overionization. The difference of $\kTz$ among ion species may result from the different initial conditions and recombination history, and from the different distributions of ion populations among different RP components.
		\item We obtain the recombination ages $6000\pm400$\,yr for the high-temperature RP and $3400\pm200$\,yr for the low-temperature RP, respectively. On the other hand, the thermal conduction timescales of the two RP components are derived to be $\sim1100$\,yr and $\sim 13000$\,yr, respectively. This indicates that the thermal conduction scenario is a possible origin for the high-temperature RP in W49B, while it may not apply to the low-temperature component. Following recent results of the hydrodynamic simulations, we suggest that different scenarios such as the adiabatic expansion and the thermal conduction may simultaneously contribute to the formation of the RP in W49B.
		\item Although the metal abundances and masses can not be perfectly characterized by existing SN models, they provide new implications and constraints on the progenitor of W49B. If W49B originates from a CC explosion, our results suggest the progenitor mass to be $\lesssim15\Msun$. But the high Mn abundance (Mn/Fe$>$1) will be confusing in the CC context. If W49B originates from a Type Ia SN, our results indicate that the metal abundance ratios {could be} roughly consistent with a DDT model with multi-spot ignition, but the X-ray emitting ejecta only account for $\sim10\%$ of the total SN ejecta.
	\end{enumerate}
	}
	\acknowledgments
	We thank Ping Zhou and Gao-Yuan Zhang for the valuable discussions.
	This work is supported by National Key R\&D Program of China No.\ 2017YFA0402600 and the NSFC under grants 11773014, 11633007, and 11851305. L. S.\ acknowledges the financial support of the China Scholar Council (No.\ 201906190108).
	\software
	{XSPEC} \citep{1996ASPC..101...17A}, SPEX \citep{1996uxsa.conf..411K}, SAS, DS9\footnote{http://ds9.si.edu/site/Home.html}

	\clearpage
	
		\begin{longrotatetable}
		\begin{deluxetable*}{p{55pt}|l|l|l|p{50pt}|p{70pt}|p{60pt}|p{70pt}|l|l|p{40pt}}
			\tablecaption{List of 15 RP-detected Galactic SNRs\label{tab:list}}
			\tablewidth{0pt}
			\tabletypesize{\scriptsize}
			\tablenum{1}
			\renewcommand\arraystretch{1.3}
			\tablehead{
				\multicolumn{5}{c}{Basic Information}\vline&\multicolumn{6}{c}{Bulk RP Property}\\ \hline
				\colhead{Name} \vline&\colhead{Age\tablenotemark{a}}\vline&\colhead{MC}\vline&\colhead{GeV}\vline&\colhead{Progenitor's }\vline&\colhead{Main Recombining}\vline&\colhead{Temperature}\vline&\colhead{Ionization Parameter}\vline&\colhead{Metal}\vline&\colhead{Possible}\vline&\colhead{Reference}\\
				\colhead{}\vline&\colhead{(kyr)}\vline&\colhead{Interaction\tablenotemark{b}}\vline&\colhead{Emission\tablenotemark{a}}\vline&\colhead{Mass\tablenotemark{c} ($M_{\odot}$)}\vline&\colhead{Feature}\vline&\colhead{$\kTe$ (keV)}\vline&\colhead{$n_{\rm e}t$ (cm$^{-3}$\,s)}\vline&\colhead{Abundance\tablenotemark{d}}\vline&\colhead{Origin of RP}\vline&\colhead{}
			}
			\startdata
			G6.4$-$0.1 (W28)&33--36&Y, OH&Y& &RRCs of He-like Si and S; Ly$\alpha$ lines of Mg and Si &LT:\,$\sim$0.1--0.2; HT:\,$\gtrsim$0.3&LT:\,$\sim$1$\times$10$^{12}$; HT:\,$\sim$(4--5)$\times$10$^{11}$&$\lesssim$&AE, TC& [1, 2, 3, 4]\\
			G31.9$+$0.0 (3C\,391)&7.4--8.4&Y, OH&Y&$\gtrsim$12&RRCs of H-like Si and S; Ly$\alpha$ lines of Si and S&$\sim$0.5&$\sim$1.4$\times$10$^{12}$&$\lesssim$&AE, TC& [5, 6]\\
			G34.7$-$0.4 (W44)&7.9--8.9&Y, OH&Y&8--15 ?& RRCs of H-like Si and S; Ly$\alpha$ line of Si&$\sim$0.4--0.5&$\sim$(6--7)$\times$10$^{11}$&$\gtrsim$&AE & [7]\\
			G43.3$-$0.2 (W49B)&2.9--6&Y ?&Y&$\gtrsim$13/$\sim$25 /SN Ia ?&RRCs of H-like Si, He-like S, and He-like Fe; Ly$\alpha$ lines of Ca and Fe &LT:\,$\sim$0.64; HT:\,$\sim$1.6 &LT:\,$\sim$5.5$\times$10$^{11}$; HT:\,$\sim$3.9$\times$10$^{11}$ &$>$ &AE, TC & This\,work; [8, 9, 10, 11, 12]\\
			G53.6$-$2.2 (3C\,400.2)&15--110& &Y& &RRCs of H-like O and Ne, and He-like Mg&$\sim$0.06--0.15 /$\sim$0.6--0.7 ?&$\sim$(1--2)$\times$10$^{11}$&$\sim$/$>$ ?& AE& [13,\,14]\\
			G89.0$+$4.7 (HB\,21)&4.8--15&Y&Y& &RRCs of H-like Si and S&$\sim$0.17&$\sim$3.2$\times$10$^{11}$&$>$& & [15]\\
			G116.9$+$0.2 (CTB\,1)&7.5--15& & &$\gtrsim$13&RRCs of H- and He-like Ne&$\sim$0.15--0.20&$\sim$(7--10.5)$\times$10$^{11}$&$\gtrsim$&TC& [16] \\
			G160.9$+$2.6 (HB\,9)&4--7&?&Y & & &$\sim$1.13&$\sim$5.8$\times$10$^{11}$&$\gtrsim$&AE& [17]\\
			G166.0$+$4.3 (VRO\,42.05.01)& &?&Y& & RRC of He-like Si; Ly$\alpha$ lines of Si and S&$\sim$0.46&$\sim$6.1$\times$10$^{11}$&$\lesssim$& TC & [18]\\
			G189.1$+$3.0 (IC\,443)&3--30&Y, OH&Y&$\gtrsim$15&He$\alpha$ lines of Fe and Ni; Ly$\alpha$ lines of Si, S, Ar, Ca and Fe; RRCs of He-like Fe and H-like Mg, Si, S and Ca &LT:\,$ \sim$0.16--0.28; HT:\,$\sim$0.48--0.67 &$\sim$5$\times$10$^{11}$ &$\gtrsim$ &AE, TC & [19, 20, 21, 22, 23]\\
			G290.1$-$0.8 (MSH\,11-61A)&10--20&?&Y &25--30&Ly$\alpha$ line of Si; RRC of H-like Si &$\sim$0.3--0.5 &$\sim$1$\times$10$^{12}$ &$>$ &AE & [24, 25] \\
			G304.6$+$0.1 (Kes\,17)&2--64& Y &Y& &Ly$\alpha$ lines of Si and S; RRC of H-like Si &$\sim$0.6--0.7 &$\sim$(1.1--1.6)$\times$10$^{12}$ &$\gtrsim$ &TC & [26]\\
			G346.6$-$0.2&4.2--16&Y, OH & & &RRCs of He-like Si and S &$\sim$0.2--0.3 &$\sim$(4--6)$\times$10$^{11}$ &$\lesssim$ &AE & [27, 28]\\
			G348.5$+$0.1 (CTB\,37A)&10--30 &Y, OH&Y & &RRCs of He-like Mg and Si &$\sim$0.5 &$\sim$1.3$\times$10$^{12}$ &$\gtrsim$ & & [29] \\
			G359.1$-$0.5& $>$10&Y, OH &Y & &Ly$\alpha$ lines of Si and S; RRCs of He-like Si and S & $\sim$0.29 & &$>$ & & [30] \\
			\enddata
			\tablecomments{The ``Y'' in the table stands for yes; the ``?'' means that the property is uncertain or controversial; the ``LT'' and ``HT'' refer to low-temperature and high-temperature component, respectively; and the ``AE'' and ``TC'' stand for adiabatic expansion and thermal conduction, respectively.}
			\tablenotetext{a}{Adopted from the {catalog} of high energy observations of Galactic SNRs \citep[][and please refer to http://www.physics.umanitoba.ca/snr/SNRcat/ for a latest online version]{2012AdSpR..49.1313F}.}
			\tablenotetext{b}{Adopted from the SNR-MC association table in \citet{2010ApJ...712.1147J}; ``OH'' means that the 1720 MHz OH maser is detected.}
			\tablenotetext{c}{Adopted from the thermal composite SNR table in \citet{2015ApJ...799..103Z}.}
			\tablenotetext{d}{Compare to 1 solar abundance.}
			\tablenotetext{}{Reference: [1] \citet{2012PASJ...64...81S}; [2] \citet{2014ApJ...791...87Z}; [3] \citet{2017ApJ...839...59P}; [4] \citet{2018PASJ...70...35O}; [5] \citet{2014ApJ...790...65E}; [6] \citet{2014PASJ...66..124S}; [7] \citet{2012PASJ...64..141U}; [8] \citet{2009ApJ...706L..71O}; [9] \citet{2010AA...514L...2M}; [10] \citet{2013ApJ...777..145L}; [11] \citet{2018AA...615A.150Z}; [12] \citet{2018ApJ...868L..35Y}; [13] \citet{2015MNRAS.446.3885B}; [14] \citet{2017ApJ...842...22E}; [15] \citet{2018PASJ...70...75S}; [16] \citet{2018PASJ...70..110K}; [17] \citet{2019arXiv190701017S}; [18] \citet{2017PASJ...69...30M}; [19] \citet{2002ApJ...572..897K}; [20] \citet{2009ApJ...705L...6Y}; [21] \citet{2014ApJ...784...74O}; [22] \citet{2017ApJ...851...73M}; [23] \citet{2018AA...615A.157G}; [24] \citet{2015PASJ...67...16K}; [25] \citet{2015ApJ...810...43A}; [26] \citet{2016PASJ...68S...4W}; [27] \citet{2013PASJ...65....6Y}; [28] \citet{2017ApJ...847..121A}; [29] \citet{2014PASJ...66....2Y}; [30] \citet{2011PASJ...63..527O}}
		\end{deluxetable*}
	\end{longrotatetable}
	
	\clearpage
	
	\begin{deluxetable*}{ll|ccc|ccc}
		\tablecaption{Observations\label{tab:obs}}
		\tabletypesize{\normalsize }
		\tablenum{2}
		\tablehead{
			\multicolumn{2}{c}{}&\multicolumn{3}{c}{$t_{\rm exp}$ (ks)\tablenotemark{a}}&\multicolumn{3}{c}{$\sum$GTI (ks)\tablenotemark{b}}\\
			ObsID&Date&MOS1&MOS2&pn&MOS1&MOS2&pn
		}
		\startdata
		0084100401&2004-04-03&18.7&18.7&17.0&17.0&16.9&14.6\\
		0084100501&2004-04-05&18.7&18.7&17.0&17.6&16.6&15.1\\
		0084100601\tablenotemark{c}&2004-04-13&2.6&2.5&-&-&-&-\\
		0724270101&2014-04-18&117.2&117.1&115.5&105.6&110.2&71.0\\
		0724270201&2014-04-19&69.9&69.8&68.3&33.8&38.9&16.8\\
		\hline
		Total& &227.1&226.8&217.8&174.0&182.6&117.5\\
		\enddata
		\tablenotetext{a}{Total exposure times.}
		\tablenotetext{b}{Total good time intervals after SP flare removal.}
		\tablenotetext{c}{Observation is heavily contaminated by SP flares.}
	\end{deluxetable*}

	\clearpage
	
	\begin{deluxetable*}{lc}
		\tablecaption{Background spectral fitting results} \label{tab:spec_bkg}
		\tablenum{3}
		\tablehead{
			Parameter&Best-fit
		}
		\startdata		
			\textbf{LHB}&${\tt apec}_{\rm LHB}$ \\
			$kT_{\rm e}$ (keV)&$0.15\pm0.01$\\
			Norm\tablenotemark{a} ($10^{-5}$\,cm$^{-5}$)&$6.48^{+1.59}_{-0.71}$\\
			\hline
			\textbf{Galactic X-ray emission} &${\tt phabs}1\times {\tt apec}_{\rm Gal}$ \\
			$N_{\rm H}$ ($10^{22}$\,cm$^{-2}$)&$3.65\pm0.04$\\
			$kT_{\rm e}$ (keV)&$1.12^{+0.01}_{-0.02}$\\
			Norm\tablenotemark{a} ($10^{-3}$\,cm$^{-5}$)&$5.66^{+0.08}_{-0.19}$\\
			\hline
			\textbf{CXB} &${\tt phabs}2\times{\tt powerlaw}_{\rm CXB}$ \\
			$N_{\rm H}$ ($10^{22}$\,cm$^{-2}$)& $=$\texttt{phabs}1$\times2$\\
			$\alpha$& 1.4 (fixed)\\
			Norm ($10^{-4}$ photons s$^{-1}$ cm$^{-2}$ keV$^{-1}$ at 1\,keV)&$1.54\pm0.07$\\
			\hline
			$\chi^2/$dof&1.18 (3378/2871)\\
		\enddata
		\tablenotetext{a}{Defined as $10^{-14}/(4\pi d^2)\int n_{\rm e} n_{\rm H} dV$.}
	\end{deluxetable*}

	\clearpage
	
	\begin{deluxetable*}{lcccc}
		\tablecaption{Spectral fitting results for global spectra of W49B}\label{tab:spec_whole}
		\tablewidth{0pt}
		\tabletypesize{\scriptsize}
		\tablenum{4}
		\tablehead{
			Parameter& \colhead{2 CIE} & \colhead{3 CIE} & \colhead{3 CIE + RRC + Gaussian} & \colhead{CIE + 2 RP}
		}
		\startdata
			\textbf{Absorption}& & & & \\
			 $N_{\rm H}$ ($10^{22}$\,cm$^{-2}$)&$8.25^{+0.04}_{-0.03}$ & $7.84\pm0.01$ &$7.37\pm0.01$ & $8.26\pm0.01$\\
			 \hline
			 \textbf{ISM} & & & & \\
			 $kT_{\rm e}$ (keV)&$0.187\pm0.001$ &$0.174\pm0.001$&$0.228\pm0.001$ &$0.177\pm0.001$\\
			 Mg&$0.30\pm0.02$ &$0.26\pm0.02$&$0.31\pm0.02$ &$0.31\pm0.02$\\
			 Norm\tablenotemark{a} (cm$^{-5}$)&$61.0^{+1.7}_{-1.4}$ &$74.0^{+0.7}_{-0.6}$&$11.0^{+0.1}_{-0.2}$ &$86.7^{+0.6}_{-0.7}$\\
			 \hline
			 \textbf{Ejecta1} & & & & \\
			 $kT_{\rm e}$ (keV)&$1.65\pm0.01$ &$1.82\pm0.01$ &$1.90^{+0.02}_{-0.01}$ &$1.60^{+0.02}_{-0.01}$ \\
			 $kT_{\rm init}$ (keV)&\nodata & \nodata &\nodata &$4.54^{+0.17}_{-0.07}$ \\
			 Si&$4.25\pm0.05$ &$4.36\pm0.04$ &$6.05^{+0.05}_{-0.07}$ & $7.61^{+0.08}_{-0.07}$\\
			 S&$4.48^{+0.04}_{-0.03}$ &$4.51^{+0.03}_{-0.04}$ &$7.50^{+0.05}_{-0.06}$ & $10.39^{+0.07}_{-0.09}$\\
			 Ar&$3.90\pm0.07$ &$4.27^{+0.07}_{-0.08}$ &$6.21^{+0.13}_{-0.11}$ & $11.46^{+0.21}_{-0.18}$\\
			 Ca&$4.12\pm0.07$ &$4.28\pm0.07$ &$6.37\pm0.11$ & $11.63^{+0.21}_{-0.17}$\\
			 Cr&$3.40\pm0.47$ &$2.43^{+0.47}_{-0.42}$ &$8.90^{+0.69}_{-0.68}$ & $13.48^{+1.02}_{-1.08}$\\
			 Mn&$2.81^{+1.02}_{-1.01}$ &$<1.52$ &$10.53^{+1.43}_{-1.44}$ & $19.40^{+2.10}_{-2.22}$\\
			 Fe&$4.99\pm0.04$ &$4.50^{+0.04}_{-0.03}$ &$7.13\pm0.06$ & $10.24^{+0.07}_{-0.11}$\\
			 $n_{\rm e}t$ ($10^{11}$\,cm$^{-3}$\,s)&\nodata &\nodata &\nodata & $3.90^{+0.08}_{-0.04}$\\
			 Redshift ($10^{-3}$)&$-2.81^{+0.02}_{-0.01}$ &$-2.79\pm0.01$ &$-2.78^{+0.01}_{-0.02}$ & $-2.79\pm0.01$\\
			 Norm\tablenotemark{a} ($10^{-2}$\,cm$^{-5}$)&$14.52^{+0.13}_{-0.15}$ &$11.96^{+0.12}_{-0.03}$ &$6.64^{+0.01}_{-0.26}$ & $5.63^{+0.04}_{-0.02}$\\
			 \hline
			 \textbf{Ejecta2} & & & & \\
			 $kT_{\rm e}$ (keV)&\nodata &$0.73\pm0.01$ &$1.02^{+0.01}_{-0.02}$ &$0.64\pm0.01$ \\
			 $kT_{\rm init}$ (keV)&\nodata &\nodata &\nodata & $2.42^{+0.05}_{-0.03}$\\
			 Abundance&\nodata & =ejecta1 &=ejecta1 & =ejecta1\\
			 $n_{\rm e}t$ ($10^{11}$\,cm$^{-3}$\,s)&\nodata &\nodata &\nodata &$5.49^{+0.04}_{-0.09}$ \\
			 Redshift ($10^{-3}$)&\nodata &$-0.28^{+0.10}_{-0.38}$ &$-1.46^{+0.08}_{-0.11}$ & $-3.74^{+0.04}_{-0.19}$\\
			 Norm\tablenotemark{a} ($10^{-2}$\,cm$^{-5}$)&\nodata &$5.31^{+0.09}_{-0.10}$ &$2.51^{+0.12}_{-0.05}$ &$10.23^{+0.10}_{-0.07}$ \\
			 \hline
			 \textbf{RRC} & & & & \\
			 Si XIV Recombining Edge (keV)&\nodata&\nodata&2.67\tablenotemark{b}&\nodata\\
			 $kT_{\rm e}$ (keV)&\nodata&\nodata&$0.64\pm0.01$&\nodata\\
			 Norm ($10^{-4}$\,photons\,s$^{-1}$\,cm$^{-2}$)&\nodata&\nodata&$6.70^{+0.33}_{-0.17}$&\nodata\\
			 S XV Recombining Edge (keV)&\nodata&\nodata&3.22\tablenotemark{b}&\nodata\\
			 $kT_{\rm e}$ (keV)&\nodata&\nodata&=Si XIV&\nodata\\
			 Norm ($10^{-4}$\,photons\,s$^{-1}$\,cm$^{-2}$)&\nodata&\nodata&$5.50^{+0.13}_{-0.16}$&\nodata\\
			 Fe XXV Recombining Edge (keV)&\nodata&\nodata&8.83\tablenotemark{b}&\nodata\\
			 $kT_{\rm e}$ (keV)&\nodata&\nodata&$1.08^{+0.20}_{-0.16}$&\nodata\\
			 Norm ($10^{-4}$\,photons\,s$^{-1}$\,cm$^{-2}$)&\nodata&\nodata&$0.84\pm0.09$&\nodata\\
			 \hline
			 \textbf{Gaussian} & & & & \\
			 Ca Ly$\alpha$ Center (keV)&\nodata&\nodata&4.10\tablenotemark{b} &\nodata\\
			 Norm ($10^{-4}$\,photons\,s$^{-1}$\,cm$^{-2}$)&\nodata&\nodata&$0.51\pm0.04$ &\nodata\\
			 Fe Ly$\alpha$ Center (keV)&\nodata&\nodata&7.00\tablenotemark{b} &\nodata\\
			 Norm ($10^{-4}$\,photons\,s$^{-1}$\,cm$^{-2}$)&\nodata&\nodata&$0.22\pm0.02$ &\nodata\\
			 \hline
			 $\chi^2$/dof &1.903\,(16069/8446) &1.775\,(14987/8443) &1.393\,(11750/8436) &1.396\,(11784/8439) \\
		\enddata
		\tablenotetext{a}{Defined as $10^{-14}/(4\pi d^2)\int n_{\rm e} n_{\rm H} dV$.}
		\tablenotetext{b}{Fixed value.}
	\end{deluxetable*}
	
	\clearpage
	
	\begin{deluxetable*}{lccc}
		\tablecaption{Energy bands used to construct EW maps.\label{tab:range}}
		\tabletypesize{\normalsize}
		\tablenum{5}
		\tablehead{
			 & \colhead{Line (eV)} & \colhead{$C_{\rm low}$ (eV)} & \colhead{$C_{\rm high}$ (eV)}
		}
		\startdata
			Si He$\alpha$ & 1770--1930 & 1500--1600 & 2750--2830 \\
			Si Ly$\alpha$ & 1930--2090 & 1500--1600 & 2750--2830 \\
			S He$\alpha$ & 2350--2550 & 1500--1600 & 2750--2830 \\
			S Ly$\alpha$ & 2550--2710 & 1500--1600 & 2750--2830 \\
			Ar He$\alpha$ & 3020--3240 & 2750--2830 & 3450--3550 \\
			Ar Ly$\alpha$ & 3240--3400 & 2750--2830 & 3450--3550 \\
			Ca He$\alpha$ & 3770--4020 & 3450--3550 & 4300--4450 \\
			Ca Ly$\alpha$ & 4020--4220 & 3450--3550 & 4300--4450 \\
			Cr & 5560--5760 & 5250--5450 & 5850--6000 \\
			Mn & 6090--6260 & 5250--5450 & 5850--6000 \\
			Fe K & 6420--6920 & 5250--5450 & 5850--6000 \\
		\enddata
	\end{deluxetable*}
	
	\clearpage
	
	\clearpage
	
	\clearpage
	
	\clearpage
	
	\clearpage
	
	\clearpage
	

	
	\clearpage
	
	\clearpage

\clearpage
	
\clearpage

\end{CJK}

\begin{thebibliography}{}
		
		\bibitem[Arnaud(1996)]{1996ASPC..101...17A} Arnaud, K.~A.\ 1996, in ASP Conf.\ Ser.\ 101, Astronomical Data Analysis Software and Systems V, ed.\ G.~H.~Jacoby \& J.~Barnes (San Francisco, CA: ASP), 17
		
		\bibitem[Auchettl et al.(2015)]{2015ApJ...810...43A} Auchettl, K., Slane, P., Castro, D., Foster, A.~R., \& Smith, R.~K.\ 2015, \apj, 810, 43
		
		\bibitem[Auchettl et al.(2017)]{2017ApJ...847..121A} Auchettl, K., Ng, C.-Y., Wong, B.~T.~T., Lopez, L., \& Slane, P.\ 2017, \apj, 847, 121
		
		\bibitem[Broersen \& Vink(2015)]{2015MNRAS.446.3885B} Broersen, S., \& Vink, J.\ 2015, \mnras, 446, 3885
		
		\bibitem[Chen et al.(2014)]{2014IAUS..296..170C} Chen, Y., Jiang, B., Zhou, P., et al.\ 2014, in IAU Symp.\ 296, Supernova Environmental Impacts, ed.\ A.~Ray \& R.~A.~McCray (Cambridge: Cambridge Univ.\ Press),  170
		
		
		\bibitem[Cowie, \& McKee(1977)]{1977ApJ...211..135C} Cowie, L.~L., \& McKee, C.~F.\ 1977, \apj, 211, 135
		
		\bibitem[Ergin et al.(2014)]{2014ApJ...790...65E} Ergin, T., Sezer, A., Saha, L., et al.\ 2014, \apj, 790, 65
		
		\bibitem[Ergin et al.(2017)]{2017ApJ...842...22E} Ergin, T., Sezer, A., Sano, H., Yamazaki, R., \& Fukui, Y.\ 2017, \apj, 842, 22 
		
		\bibitem[Ferrand \& Safi-Harb(2012)]{2012AdSpR..49.1313F} Ferrand, G., \& Safi-Harb, S.\ 2012, Advances in Space Research, 49, 1313
		
		\bibitem[Greco et al.(2018)]{2018AA...615A.157G} Greco, E., Miceli, M., Orlando, S., et al.\ 2018, \aap, 615, A157
		
		\bibitem[H.~E.~S.~S. Collaboration et al.(2018)]{2018A&A...612A...5H} H.~E.~S.~S. Collaboration, Abdalla, H., Abramowski, A., et al.\ 2018, \aap, 612, A5
		
		\bibitem[Hitomi Collaboration et al.(2018)]{2018PASJ...70...12H} Hitomi Collaboration, Aharonian, F., Akamatsu, H., et al.\ 2018, \pasj, 70, 12
		
		\bibitem[Hwang et al.(2000)]{2000ApJ...532..970H} Hwang, U., Petre, R., \& Hughes, J.~P.\ 2000, \apj, 532, 970
		
		\bibitem[Hwang et al.(2000)]{2000ApJ...537L.119H} Hwang, U., Holt, S.~S., \& Petre, R.\ 2000, \apjl, 537, L119
		
		\bibitem[Itoh \& Masai(1989)]{1989MNRAS.236..885I} Itoh, H., \& Masai, K.\ 1989, \mnras, 236, 885
		
		\bibitem[Jiang et al.(2010)]{2010ApJ...712.1147J} Jiang, B., Chen, Y., Wang, J., et al.\ 2010, \apj, 712, 1147
		
		\bibitem[Kaastra et al.(1996)]{1996uxsa.conf..411K} Kaastra, J.~S., Mewe, R., \& Nieuwenhuijzen, H.\ 1996, in 11th Coll.\ on UV and X-ray Spectroscopy of Astrophysical and Laboratory Plasmas, ed.\ K. Yamashita \& T. Watanabe (Tokyo: Universal Academy Press), 411
		
		\bibitem[Kamitsukasa et al.(2015)]{2015PASJ...67...16K} Kamitsukasa, F., Koyama, K., Uchida, H., et al.\ 2015, \pasj, 67, 16
		
		\bibitem[Katayama et al.(2004)]{2004A&A...414..767K} Katayama, H., Takahashi, I., Ikebe, Y., et al.\ 2004, \aap, 414, 767
		
		\bibitem[Katsuragawa et al.(2018)]{2018PASJ...70..110K} Katsuragawa, M., Nakashima, S., Matsumura, H., et al.\ 2018, \pasj, 70, 110
		
		\bibitem[Kawasaki et al.(2002)]{2002ApJ...572..897K} Kawasaki, M.~T., Ozaki, M., Nagase, F., et al.\ 2002, \apj, 572, 897
		
		\bibitem[Kawasaki et al.(2005)]{2005ApJ...631..935K} Kawasaki, M., Ozaki, M., Nagase, F., et al.\ 2005, \apj, 631, 935
		
		\bibitem[Keohane et al.(2007)]{2007ApJ...654..938K} Keohane, J.~W., Reach, W.~T., Rho, J., et al.\ 2007, \apj, 654, 938
		
		
		\bibitem[Kuntz, \& Snowden(2008)]{2008A&A...478..575K} Kuntz, K.~D., \& Snowden, S.~L.\ 2008, \aap, 478, 575
		
		\bibitem[Lopez et al.(2009)]{2009ApJ...706L.106L} Lopez, L.~A., Ramirez-Ruiz, E., Badenes, C., et al.\ 2009, \apjl, 706, L106
		
		\bibitem[Lopez et al.(2011)]{2011ApJ...732..114L} Lopez, L.~A., Ramirez-Ruiz, E., Huppenkothen, D., et al.\ 2011, \apj, 732, 114
		
		\bibitem[Lopez et al.(2013)]{2013ApJ...764...50L} Lopez, L.~A., Ramirez-Ruiz, E., Castro, D., \& Pearson, S.\ 2013, \apj, 764, 50
		
		\bibitem[Lopez et al.(2013)]{2013ApJ...777..145L} Lopez, L.~A., Pearson, S., Ramirez-Ruiz, E., et al.\ 2013, \apj, 777, 145 
		
		\bibitem[Maeda et al.(2010)]{2010ApJ...712..624M} Maeda, K., R{\"o}pke, F.~K., Fink, M., et al.\ 2010, \apj, 712, 624
		
		\bibitem[Masui et al.(2009)]{2009PASJ...61S.115M} Masui, K., Mitsuda, K., Yamasaki, N.~Y., et al.\ 2009, \pasj, 61, S115 
		
		\bibitem[Matsumura et al.(2017)]{2017PASJ...69...30M} Matsumura, H., Uchida, H., Tanaka, T., et al.\ 2017, \pasj, 69, 30 
		
		\bibitem[Matsumura et al.(2017)]{2017ApJ...851...73M} Matsumura, H., Tanaka, T., Uchida, H., Okon, H., \& Tsuru, T.~G.\ 2017, \apj, 851, 73
		
		\bibitem[Mernier et al.(2019)]{2019arXiv191109684M} Mernier, F., Werner, N., Lakhchaura, K., et al.\ 2019, arXiv e-prints, arXiv:1911.09684
		
		\bibitem[Miceli et al.(2006)]{2006A&A...453..567M} Miceli, M., Decourchelle, A., Ballet, J., et al.\ 2006, \aap, 453, 567 
		
		\bibitem[Miceli et al.(2010)]{2010AA...514L...2M} Miceli, M., Bocchino, F., Decourchelle, A., Ballet, J., \& Reale, F.\ 2010, \aap, 514, L2
		
		\bibitem[Moffett, \& Reynolds(1994)]{1994ApJ...437..705M} Moffett, D.~A., \& Reynolds, S.~P.\ 1994, \apj, 437, 705
		
		\bibitem[Nomoto et al.(2006)]{2006NuPhA.777..424N} Nomoto, K., Tominaga, N., Umeda, H., Kobayashi, C., \& Maeda, K.\ 2006, Nuclear Physics A, 777, 424 
		
		\bibitem[Nomoto et al.(2013)]{2013ARA&A..51..457N} Nomoto, K., Kobayashi, C., \& Tominaga, N.\ 2013, \araa, 51, 457 
		
		\bibitem[Ohnishi et al.(2011)]{2011PASJ...63..527O} Ohnishi, T., Koyama, K., Tsuru, T.~G., et al.\ 2011, \pasj, 63, 527
		
		\bibitem[Ohnishi et al.(2014)]{2014ApJ...784...74O} Ohnishi, T., Uchida, H., Tsuru, T.~G., et al.\ 2014, \apj, 784, 74
		
		\bibitem[Okon et al.(2018)]{2018PASJ...70...35O} Okon, H., Uchida, H., Tanaka, T., Matsumura, H., \& Tsuru, T.~G.\ 2018, \pasj, 70, 35
		
		\bibitem[Orlando et al.(2005)]{2005A&A...444..505O} Orlando, S., Peres, G., Reale, F., et al.\ 2005, \aap, 444, 505
		
		\bibitem[Ozawa et al.(2009)]{2009ApJ...706L..71O} Ozawa, M., Koyama, K., Yamaguchi, H., Masai, K., \& Tamagawa, T.\ 2009, \apjl, 706, L71 
		
		\bibitem[Pannuti et al.(2017)]{2017ApJ...839...59P} Pannuti, T.~G., Rho, J., Kargaltsev, O., et al.\ 2017, \apj, 839, 59
		
		\bibitem[Patnaude et al.(2015)]{2015ApJ...803..101P} Patnaude, D.~J., Lee, S.-H., Slane, P.~O., et al.\ 2015, \apj, 803, 101
		
		\bibitem[Patnaude \& Badenes(2017)]{2017hsn..book.2233P} Patnaude, D., \& Badenes, C.\ 2017, in Handbook of Supernovae, ed.\ A.~W.~Alsabti \& P.~Murdin (Berlin: Springer), 2233
		
		\bibitem[Pye et al.(1984)]{1984MNRAS.207..649P} Pye, J.~P., Becker, R.~H., Seward, F.~D., et al.\ 1984, \mnras, 207, 649
		
		\bibitem[Radhakrishnan et al.(1972)]{1972ApJS...24...49R} Radhakrishnan, V., Goss, W.~M., Murray, J.~D., et al.\ 1972, \apjs, 24, 49
		
		\bibitem[Ranasinghe, \& Leahy(2018)]{2018AJ....155..204R} Ranasinghe, S., \& Leahy, D.~A.\ 2018, \aj, 155, 204
		
		\bibitem[Rho, \& Petre(1998)]{1998ApJ...503L.167R} Rho, J., \& Petre, R.\ 1998, \apjl, 503, L167
		
		\bibitem[Sato et al.(2014)]{2014PASJ...66..124S} Sato, T., Koyama, K., Takahashi, T., Odaka, H., \& Nakashima, S.\ 2014, \pasj, 66, 124
		
		\bibitem[Sawada \& Koyama(2012)]{2012PASJ...64...81S} Sawada, M., \& Koyama, K.\ 2012, \pasj, 64, 81
		
		\bibitem[Seitenzahl et al.(2013)]{2013MNRAS.429.1156S} Seitenzahl, I.~R., Ciaraldi-Schoolmann, F., R{\"o}pke, F.~K., et al.\ 2013, \mnras, 429, 1156
		
		\bibitem[Sezer et al.(2019)]{2019arXiv190701017S} Sezer, A., Ergin, T., Yamazaki, R., et al.\ 2019, arXiv e-prints, arXiv:1907.01017
		
		
		\bibitem[Smith et al.(1985)]{1985ApJ...296..469S} Smith, A., Jones, L.~R., Peacock, A., et al.\ 1985, \apj, 296, 469
		
		\bibitem[Smith \& Hughes(2010)]{2010ApJ...718..583S} Smith, R.~K., \& Hughes, J.~P.\ 2010, \apj, 718, 583
		
		\bibitem[Spitzer(1962)]{1962pfig.book.....S} Spitzer, L.\ 1962, Physics of Fully Ionized Gases (New York: Interscience)
		
		\bibitem[Sukhbold et al.(2016)]{2016ApJ...821...38S} Sukhbold, T., Ertl, T., Woosley, S.~E., Brown, J.~M., \& Janka, H.-T.\ 2016, \apj, 821, 38
		
		\bibitem[Sun, \& Chen(2019)]{2019ApJ...872...45S} Sun, L., \& Chen, Y.\ 2019, \apj, 872, 45
		
		\bibitem[Suzuki et al.(2018)]{2018PASJ...70...75S} Suzuki, H., Bamba, A., Nakazawa, K., et al.\ 2018, \pasj, 70, 75 
		
		
		\bibitem[Thielemann et al.(2018)]{2018SSRv..214...62T} Thielemann, F.-K., Isern, J., Perego, A., \& von Ballmoos, P.\ 2018, \ssr, 214, 62 
		
		\bibitem[Uchida et al.(2012)]{2012PASJ...64..141U} Uchida, H., Koyama, K., Yamaguchi, H., et al.\ 2012, \pasj, 64, 141 
		
		\bibitem[Uchiyama et al.(2013)]{2013PASJ...65...19U} Uchiyama, H., Nobukawa, M., Tsuru, T.~G., \& Koyama, K.\ 2013, \pasj, 65, 19
		
		\bibitem[Washino et al.(2016)]{2016PASJ...68S...4W} Washino, R., Uchida, H., Nobukawa, M., et al.\ 2016, \pasj, 68, S4 
		
		\bibitem[Yamaguchi et al.(2009)]{2009ApJ...705L...6Y} Yamaguchi, H., Ozawa, M., Koyama, K., et al.\ 2009, \apjl, 705, L6 
		
		\bibitem[Yamaguchi et al.(2014)]{2014ApJ...785L..27Y} Yamaguchi, H., Badenes, C., Petre, R., et al.\ 2014, \apjl, 785, L27
		
		\bibitem[Yamaguchi et al.(2018)]{2018ApJ...868L..35Y} Yamaguchi, H., Tanaka, T., Wik, D.~R., et al.\ 2018, \apjl, 868, L35 
		
		\bibitem[Yamauchi et al.(2013)]{2013PASJ...65....6Y} Yamauchi, S., Nobukawa, M., Koyama, K., \& Yonemori, M.\ 2013, \pasj, 65, 6
		
		\bibitem[Yamauchi et al.(2014)]{2014PASJ...66....2Y} Yamauchi, S., Minami, S., Ota, N., \& Koyama, K.\ 2014, \pasj, 66, 2
		
		\bibitem[Yang et al.(2013)]{2013ApJ...766...44Y} Yang, X.~J., Tsunemi, H., Lu, F.~J., et al.\ 2013, \apj, 766, 44
		
		\bibitem[Zhang et al.(2015)]{2015ApJ...799..103Z} Zhang, G.-Y., Chen, Y., Su, Y., et al.\ 2015, \apj, 799, 103
		
		\bibitem[Zhang et al.(2019)]{2019ApJ...875...81Z} Zhang, G.-Y., Slavin, J.~D., Foster, A., et al.\ 2019, \apj, 875, 81 
		
		\bibitem[Zhou et al.(2014)]{2014ApJ...791...87Z} Zhou, P., Safi-Harb, S., Chen, Y., et al.\ 2014, \apj, 791, 87
		
		\bibitem[Zhou \& Vink(2018)]{2018AA...615A.150Z} Zhou, P., \& Vink, J.\ 2018, \aap, 615, A150
		
		\bibitem[Zhou et al.(2011)]{2011MNRAS.415..244Z} Zhou, X., Miceli, M., Bocchino, F., Orlando, S., \& Chen, Y.\ 2011, \mnras, 415, 244 
		
		\bibitem[Zhu et al.(2014)]{2014ApJ...793...95Z} Zhu, H., Tian, W.~W., \& Zuo, P.\ 2014, \apj, 793, 95 
		
		
	\end{thebibliography}
\end{document}